\documentclass[10pt,twocolumn]{article}

\usepackage[utf8]{inputenc}
\usepackage[T1]{fontenc}
\usepackage{amsmath,amssymb}
\usepackage{graphicx}
\usepackage{booktabs}
\usepackage{hyperref}
\usepackage{listings}
\usepackage{xcolor}
\usepackage[margin=0.85in]{geometry}
\usepackage{enumitem}
\usepackage{caption}
\usepackage{subcaption}
\usepackage{algorithm}
\usepackage{algpseudocode}
\usepackage{multirow}
\usepackage{tabularx}
\usepackage{tikz}
\usetikzlibrary{positioning,arrows.meta,shapes.geometric,fit,backgrounds,calc}

\title{\textbf{Themis: A Filesystem Model Checker\\That Owns the Machine}}

\author{
  Daeyeon Son\\
  Independent Researcher\\
  Republic of Korea\\
  \texttt{sdy1350@gmail.com}
}

\date{August 2026}

\begin{document}
\maketitle

% ============================================================
\begin{abstract}
File-system model checkers explore an unmodified in-kernel filesystem's
state space to find bugs that escape unit tests. The state of the art,
Metis~\cite{metis-fast24}, runs \emph{inside} the operating system. It
drives syscalls and reads their returns. Lacking a cheap way to snapshot
in-kernel state from user space, it hand-codes a well-engineered reference
filesystem, RefFS, to serve as its differential oracle. Where a checker
sits shapes what it can see. Metis sits above the block device and above
the clock: it observes the filesystem's syscall interface rather than the
device beneath it, and it sets time aside from its abstract state as noise.
Both are documented choices.

I present \textbf{Themis}, a filesystem model checker built \emph{beneath}
an unmodified in-kernel filesystem on a bare-metal Type-1 AMD-V hypervisor.
Themis \textbf{owns the machine}: it owns the virtual block device and the
virtual clock underneath a stock Linux guest running a stock ext2 driver.
Ownership buys four things. First, a \textbf{machine-layer
$\varepsilon$-copy fork}: a VM/VMCB snapshot, nested-page-table
write-protection of the guest working set, and a copy-on-write disk
overlay. Its restore cost is proportional to \emph{dirtied pages, not
filesystem image size}, so \textbf{every unmodified in-kernel filesystem
becomes its own differential reference and a hand-coded reference is no
longer required}. Second, \textbf{machine-owned observation}: per-operation
below-FS write/read/location profiles, and byte-level whole-disk state diffs
with nuisance-field normalization. Third, \textbf{control of the clock},
which turns time into an explored search dimension. Fourth, the same
$\varepsilon$-copy fork doubles as a re-armable checkpoint for bit-exact
\textbf{deterministic replay}: a real ext2 workload replays whole-DRAM and
whole-disk bit-exact across nine passes, and a timing-dependent outcome that
is flaky raw (nine distinct hashes) becomes reproducible when served (nine
identical). This replay machinery is so far exercised on synthetic workloads
rather than the filesystem headline.

I report a silicon campaign of over 40 hardware runs across four days on an
AMD Zen5 9800X3D. The mechanism is proven end-to-end: a real virtio-blk filesystem
write is served \emph{inside} a fork window and then fully reverted---DRAM
(a whole-working-set hash matches byte-for-byte), disk (a host-side overlay
oracle reverts), and device state---in \textbf{9\,$\mu$s for the dirtied
set} (O(dirtied)).

The principal result is the clock. Themis drives a real ext2 filesystem
across a clock sweep and witnesses a real Y2038 defect on silicon. A
128-byte-inode ext2 clamps any post-2038 timestamp to \texttt{0x7FFFFFFF},
and the clamp persists across a remount---a property of the 128-byte inode,
not of ext2: ext4 with 128-byte inodes clamps identically, and ext2 with
256-byte inodes does not. A byte-level whole-disk diff
between two logically identical runs---after canonicalizing the filesystem's
random-identity fields (\texttt{s\_uuid}, \texttt{s\_hash\_seed},
\texttt{i\_generation}) out---goes from three differing sectors to
\textbf{zero}: the disk is deterministic \emph{modulo nuisance}. Because the
defect lives in the timestamps, a time-excluded fingerprint of the Metis
kind does not register it; owning the clock is what makes it visible.

Two further silicon results, both on real ext2, support that thesis. First,
below-filesystem corruption microscopy: flipping a single bit in a file's
data block is read back wrong and reported by \emph{zero} filesystem error
(ext2 has no data checksums---invisible to any in-OS checker), while a
single-bit flip of the superblock magic renders the whole filesystem
unmountable---a demonstrated blast-radius spectrum from silent to fatal.
Second, throughput, which I frame honestly as a backtrack \textbf{cost
model} rather than a states/second count: Themis reverts only dirtied pages
while Metis reloads the whole image, a gap that \emph{widens} with image
size (165$\times$ vs Metis's ext4 remount, 2100$\times$ vs xfs), reported
this way (a revert-only backtrack cost, not end-to-end throughput) because I
estimate my bare-metal guest runs roughly an order of magnitude slower. I
deliberately \textbf{do not} claim crash-consistency (a crowded corner I
concede), and I \textbf{drop} attestation as a headline; I position both as
future cross-product work realizing the Metis group's own dissertation
\S7.1.1.
\end{abstract}

% ============================================================
\section{Introduction}
\label{sec:intro}

A filesystem is a large, concurrent, crash-exposed state machine, and the
bugs that matter most are the ones that survive years of production because
no test drives the filesystem into the corner that triggers them. Model
checking attacks this by systematically exploring the reachable states:
from a state, apply each candidate operation, canonicalize the result into
an abstract-state key, deduplicate against the visited set, and backtrack.
The FAST'24 system Metis~\cite{metis-fast24} showed this is practical
against \emph{unmodified in-kernel} filesystems and found real, long-lived
bugs (including a 16-year-old JFFS2 defect). Metis pioneered versatile,
in-OS filesystem model checking, and running inside the OS is a sound
choice: the vantage is simple, portable, and fast, and it produced real
results against real, unmodified filesystems.

Metis makes two design choices that follow from its architecture, both
stated in its own paper. First, running inside the OS as a userspace
process, it does not have a cheap way to snapshot and restore the
\emph{in-kernel} state of the filesystem under test---the page cache, the
inode cache, the journal, the block device. Metis reports that it ``tried
and evaluated \ldots process snapshotting, VM snapshotting, and LightVM. None
\ldots were effective due to functional deficiencies or inefficient
performance''~\cite{metis-fast24}, and, having found no suitable
off-the-shelf option, it builds \textbf{RefFS}, a purpose-built RAM-FUSE
reference filesystem ($\sim$4{,}000 LoC) whose purpose is to be
snapshottable, to serve as the differential oracle. Second, Metis
\textbf{sets time aside} in its abstract state---atime, mtime, ctime are
``noisy attributes''---and its \S3.7 notes it ``cannot detect atime-related
bugs.'' Both choices follow from \emph{where Metis sits}: above the block
device and above the clock.

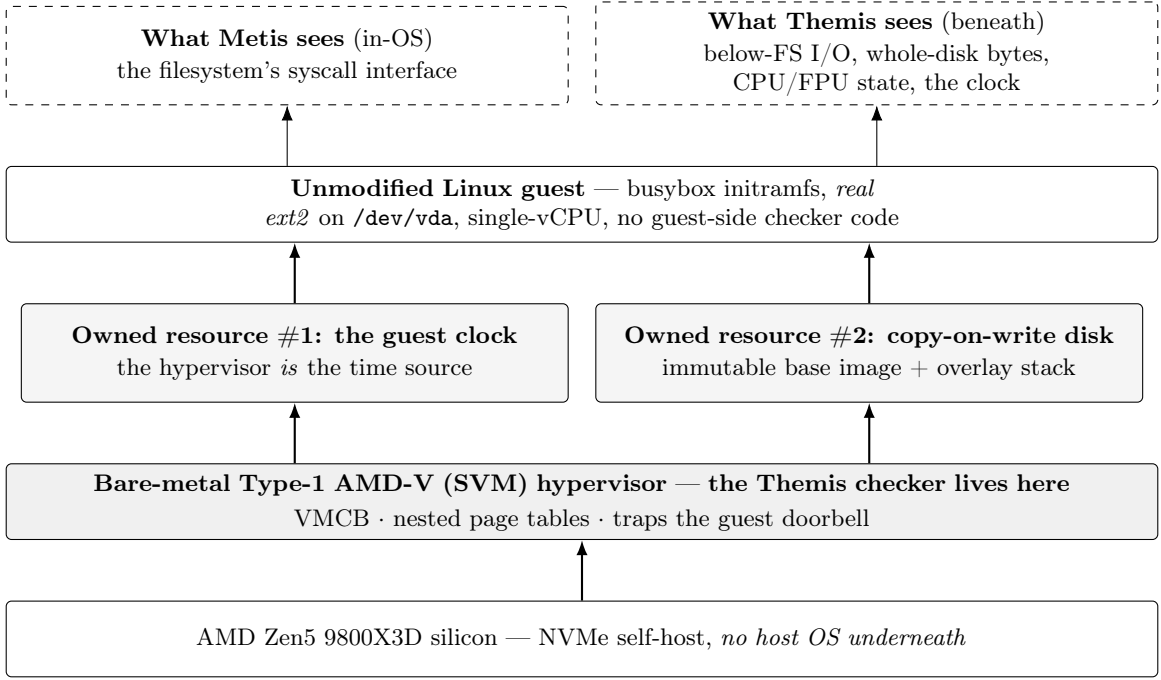
\begin{figure*}[t]\centering
\begin{tikzpicture}[
  font=\small, node distance=8mm,
  layer/.style={draw, rounded corners=2pt, align=center, minimum height=10mm, text width=150mm},
  own/.style={draw, rounded corners=2pt, align=center, minimum height=13mm, text width=70mm, fill=black!4},
  vis/.style={draw, dashed, rounded corners=2pt, align=center, minimum height=13mm, text width=72mm},
  owns/.style={-{Latex[length=2mm]}, thick},
  >={Latex[length=2mm]}]

  \node[layer] (si)
    {AMD Zen5 9800X3D silicon --- NVMe self-host, \emph{no host OS underneath}};
  \node[layer, above=8mm of si, fill=black!6] (hv)
    {\textbf{Bare-metal Type-1 AMD-V (SVM) hypervisor} --- \textbf{the Themis checker lives here}\\[2pt]
     VMCB $\cdot$ nested page tables $\cdot$ traps the guest doorbell};
  \node[own, above=8mm of hv, xshift=-38mm] (clk)
    {\textbf{Owned resource \#1: the guest clock}\\[1pt] the hypervisor \emph{is} the time source};
  \node[own, above=8mm of hv, xshift=38mm] (blk)
    {\textbf{Owned resource \#2: copy-on-write disk}\\[1pt] immutable base image + overlay stack};
  \node[layer, anchor=south] (guest) at ([yshift=8mm]clk.north -| si)
    {\textbf{Unmodified Linux guest} --- busybox initramfs, \emph{real ext2} on
     \texttt{/dev/vda}, single-vCPU, no guest-side checker code};

  \draw[owns] (si.north) -- (hv.south);
  \draw[owns] (hv.north -| clk.south) -- (clk.south);
  \draw[owns] (hv.north -| blk.south) -- (blk.south);
  \draw[owns] (clk.north) -- (clk.north |- guest.south);
  \draw[owns] (blk.north) -- (blk.north |- guest.south);

  \node[vis, above=8mm of guest.north west, anchor=south west] (mv)
    {\textbf{What Metis sees} (in-OS)\\[1pt] the filesystem's syscall interface};
  \node[vis, above=8mm of guest.north east, anchor=south east] (tv)
    {\textbf{What Themis sees} (beneath)\\[1pt] below-FS I/O, whole-disk bytes, CPU/FPU state, the clock};
  \draw[->] (mv.south |- guest.north) -- (mv.south);
  \draw[->] (tv.south |- guest.north) -- (tv.south);
\end{tikzpicture}
\caption{\textbf{The ownership cascade.} A bare-metal Type-1 AMD-V
  hypervisor---where the Themis checker lives---owns the guest's
  copy-on-write block device and clock beneath an unmodified Linux guest
  that runs a stock ext2 driver with no guest-side checker code. The dashed
  columns contrast visibility: Metis, inside the OS, observes the
  syscall-return interface; Themis, beneath the guest, also observes below-FS
  I/O, whole-disk bytes, and CPU/FPU state.}
\label{fig:cascade}
\end{figure*}

The idea is to move the checker \emph{beneath} the filesystem. If the checker owns the
machine---a Type-1 hypervisor that owns the virtual block device and the
virtual clock under an unmodified in-kernel filesystem---then the two forced
choices dissolve (Figure~\ref{fig:cascade}):

\begin{itemize}[topsep=2pt,itemsep=2pt,leftmargin=1.2em]
  \item \textbf{Own the block device $\Rightarrow$ every filesystem is its
    own reference.} A machine-layer fork can snapshot and restore the
    \emph{entire} in-kernel filesystem state cheaply, so an \emph{unmodified}
    ext2/ext4/xfs can play the role RefFS was invented to play. Two forked
    instances (or one instance before and after a transformation) driven
    through the same operations are a differential oracle with no hand-built
    reference---an alternative route to the cheap snapshotting RefFS
    provides, reached by making the real filesystem itself cheap to restore.
  \item \textbf{Own the block device $\Rightarrow$ observe and inject below
    the filesystem.} The checker measures, per operation, exactly how many
    sectors the filesystem wrote, how many it read, and \emph{where}---
    reverse-engineering the on-disk layout from beneath, and diffing
    whole-disk state at byte granularity. It can flip a single bit in a
    chosen on-disk structure and watch the blast radius. All of this follows
    from sitting beneath the driver rather than at the syscall boundary.
  \item \textbf{Own the clock $\Rightarrow$ time becomes a search axis.} The
    checker drives the virtual wall clock through adversarial values (the
    2038 overflow, rollback, near-equality) and folds the resulting on-disk
    timestamps into the abstract state---exactly the dimension Metis sets
    aside as noise.
\end{itemize}

Of these three levers, the clock is this paper's headline. It is the search
axis an in-OS checker sets aside as noise, and it is where Themis witnesses a
real Y2038 defect (\S\ref{sec:e1}). The below-filesystem observation and the
deterministic replay support that result rather than share the billing.

The honest core of the argument is this. The fork itself is not a
contribution: lightweight VM snapshot/restore is
commodity~\cite{potemkin,snowflock,remus,firecracker,xenfork,agamotto,nyx}.
The contribution is what owning the machine \emph{enables}. Above all it
enables control of the clock; it also removes the need for a hand-coded
reference and lets the checker see below the filesystem. And it demonstrates
that a machine-layer fork serves, from beneath the OS, the snapshotting task
the FAST'24 baseline evaluated and, within its in-OS architecture, found
unsuitable. I am equally explicit about what this paper does \textbf{not}
claim. It does
\textbf{not} do crash-consistency testing---a crowded, conceded
corner~\cite{crashmonkey,dmlogwrites,alice,explode,fisc,hydra,chipmunk,vinter,snapcc}
where owning the machine adds little over an in-kernel bio log; I frame the
crash $\times$ clock $\times$ corruption cross-product as future work that
realizes the Metis group's own dissertation
\S7.1.1~\cite{yifei-dissertation}. It does \textbf{not} lead with
attestation; the hash-chained write log is quiet infrastructure, pre-empted
as a novelty by PeerReview~\cite{peerreview}. And it does \textbf{not} yet
report a \emph{previously-unknown} reproduced bug---the Y2038 and
coarse-timestamp results are real but known classes; the new-bug hunt is
future work.

This paper makes the following contributions.
\begin{enumerate}[topsep=2pt,itemsep=2pt,leftmargin=1.3em]
  \item \textbf{A machine-layer $\varepsilon$-copy fork as the state-restore
    primitive for filesystem model checking.} A whole-VMCB + GPR + XSAVE +
    device snapshot, nested-page-table write-protection of the guest working
    set, and a copy-on-write disk overlay, with restore in time
    \textbf{O(dirtied pages)} and O(1) backtracking. Silicon-proven
    end-to-end and reverting in \textbf{9\,$\mu$s} for the dirtied set
    (\S\ref{sec:eval}). This is what lets every unmodified in-kernel
    filesystem be its own reference, removing the need for a hand-coded one,
    and it realizes from beneath the OS the snapshotting Metis evaluated and
    set aside as unsuitable for an in-OS checker.
  \item \textbf{Machine-owned below-filesystem observation.} Per-operation
    write-amplification, write-localization (from which I reverse-engineer
    the ext2 on-disk layout without reading a line of filesystem source),
    and read-amplification profiles; and a byte-level whole-disk diff with
    \textbf{nuisance-field normalization} (\S\ref{sec:observe}).
  \item \textbf{The clock as a first-class explored dimension (the principal result).}
    A differential clock sweep over a real filesystem that catches a real
    Y2038 timestamp defect that a time-excluded (Metis-parity) fingerprint
    does not register (\S\ref{sec:e1}).
  \item \textbf{Causal below-filesystem corruption microscopy.} A
    single-bit flip beneath the filesystem whose blast radius spans a
    demonstrated spectrum from \emph{silent} to \emph{fatal}
    (\S\ref{sec:e2}).
  \item \textbf{A backtrack cost model, honestly framed.} Themis reverts
    O(dirtied); Metis reloads O(image). The gap \emph{widens} with
    filesystem size---165$\times$ vs Metis's published ext4 remount,
    2100$\times$ vs xfs---presented as a revert-only cost model, not a
    states/second number, because I estimate my bare-metal guest runs
    roughly an order of magnitude slower (\S\ref{sec:eval}).
  \item \textbf{Deterministic replay from the same re-armable fork.} The
    machine-owned $\varepsilon$-copy fork doubles as a re-armable checkpoint
    for bit-exact deterministic replay: a real ext2 create+write+fsync
    workload replays whole-DRAM and whole-disk bit-exact across nine passes,
    and a timing-dependent outcome that is flaky with the raw counter (nine
    distinct hashes) becomes reproducible when the counter is served (nine
    identical) (\S\ref{sec:replay}).
\end{enumerate}

The system is one component of AnimaOS, a bare-metal Rust operating system
that includes the Type-1 AMD-V hypervisor Themis is built on. All silicon
results are from a campaign of over 40 hardware runs across four days (July
13, 14, 15, and 18) on an AMD Zen5 9800X3D self-hosting from NVMe; bootlogs
are archived.

% ============================================================
\section{Background and Related Work}
\label{sec:related}

\label{sec:metis}
Metis~\cite{metis-fast24} is the immediate predecessor, and the system
whose exploration approach Themis reuses and extends. Metis pioneered
versatile in-OS filesystem model checking; its in-OS vantage is simple,
portable, and fast, and its design choices---RefFS and setting time
aside---are reasonable engineering within that architecture. Themis
deliberately takes the opposite architectural bet, moving the checker
beneath the OS, and inherits a different set of trade-offs: more machine
control at the cost of a heavier substrate and a slower guest. The two are
complementary explorations of the same design space. Metis's
structure---which Themis reuses verbatim in its abstract-state policy---is: a
fixed set of filesystem operations (five metadata operations plus ten
syscalls), argument binning into power-of-two partitions with
per-argument-class distributions, and a bounded depth-first search (default
depth 10{,}000, reduced for large-device filesystems) with an abstract-state
fingerprint for deduplication. Metis drives the real, unmodified in-kernel
filesystem via syscalls and compares it against a differential reference. Its
two design choices, both stated in the paper and both reasonable within its
architecture, are the points where Themis's beneath-the-OS vantage takes a
different route:

\begin{itemize}[topsep=2pt,itemsep=2pt,leftmargin=1.2em]
  \item \textbf{RefFS.} Metis needs a snapshottable reference. Having
    evaluated process snapshotting, VM snapshotting, and LightVM and found
    them ``not effective due to functional deficiencies or inefficient
    performance''~\cite{metis-fast24}, it implements RefFS, a
    $\sim$4{,}000-LoC RAM-FUSE filesystem with ioctl
    SAVE/RESTORE/PICKLE/LOAD ($\sim$12.5\,KB per state), purpose-built to be
    cheap to snapshot. RefFS is a well-engineered solution to a real
    problem---cheap snapshotting under an in-OS checker; Themis reaches the
    same goal by a different route, making the real filesystem itself cheap
    to restore, so a hand-coded reference is not required.
  \item \textbf{Time set aside.} Metis's abstract state excludes
    atime/mtime/ctime as noisy attributes---a reasonable simplification for a
    checker whose vantage does not read the on-disk timestamp path---and its
    \S3.7 notes this leaves atime-related bugs out of scope. Time is the
    dimension Metis sets aside as noise, which Themis's clock chapter
    re-examines.
\end{itemize}

Metis's published throughput---RefFS $\sim$830 ops/s and $\sim$349.9
states/s, ext4 $\sim$280.3 ops/s and $\sim$112.6 states/s, xfs $\sim$29.2
ops/s and $\sim$8.8 states/s~\cite{metis-fast24}---is the baseline for my
backtrack cost model (\S\ref{sec:eval}). The two-order-of-magnitude gap
between RefFS and xfs is precisely the \emph{O(image) reload} cost of using
a real filesystem as its own reference in an in-OS checker---the cost
Themis's O(dirtied) fork removes. EXPLODE and FiSC~\cite{explode,fisc} are the
model-checking ancestors and are worth positioning against precisely, because
Themis inherits their core idea and moves only the substrate. FiSC
model-checked Linux filesystems from inside a modified kernel, and EXPLODE
generalized it to a lightweight, in-situ storage checker that snapshots and
restores the running system to explore states and found serious errors across
a wide range of storage stacks. Both took the exploration approach Themis
reuses; both, however, did their checkpointing \emph{inside} the machine they
were testing, and neither observed below the filesystem or drove the clock.
FiSC explicitly notes that ``any VMM is a valid substitute'' for its in-kernel
checkpointing---a line that both anticipates and licenses the Themis
substrate. Themis takes that license literally: it keeps the exploration
engine and pushes the checkpoint down to a machine-layer fork, which is
exactly what opens the below-FS and clock dimensions those ancestors left
closed.

\label{sec:crash}
I turn next to crash-consistency testing, positioning against that
literature \emph{proactively} so
that no reviewer mistakes Themis for a crash tester. Crash-consistency is a
mature, crowded area: CrashMonkey/ACE/B3~\cite{crashmonkey} systematically
generate small workloads (ACE) and bounded-exhaustively test recovery from
the crash states each one can reach (the B3 approach), catching real
crash-consistency bugs in mature filesystems; \texttt{dm-log-writes}~\cite{dmlogwrites}
is an in-tree device-mapper target that records every bio with FLUSH/FUA
ordering and replays barrier-respecting prefixes under an \emph{unmodified}
filesystem; BOB/ALICE~\cite{alice} reorder and drop writes at barriers to
find application-level crash vulnerabilities against an abstract model of
what each filesystem persists; Hydra~\cite{hydra} and Chipmunk~\cite{chipmunk}
scale coverage; Vinter~\cite{vinter} does full-system crash testing of an
unmodified kernel; and SnapCC~\cite{snapcc} pairs a fuzzer with per-workload
crash states.

The critical point for Themis: on the crash axis, ``owning the machine''
adds essentially nothing. \texttt{dm-log-writes} already sees every bio and
every barrier beneath an unmodified filesystem, so a ``hypervisor that logs
writes and replays prefixes'' is a re-implementation of an in-tree tool one
layer down. Two adversarial design reviews independently reached a unanimous
no-go on shipping crash consistency in this paper. \textbf{Themis does not
implement crash-consistency testing.} The only non-derivative framing---
per-node-of-the-\emph{global}-model-checker-graph crash enumeration, with
the fork/replay substrate making it cheap and sound---is exactly the
integration the Metis group's own PhD dissertation
\S7.1.1~\cite{yifei-dissertation} lists as future work (``crash state must
be considered as part of the overall state description''). I claim only to
\emph{realize their open problem} as future cross-product work
(\S\ref{sec:limitations}), not to compete with the crash-testing crowd here.

\label{sec:corruptfake}
On corruption injection and clock faking, the closest relative to the
corruption experiments (below-FS corruption) is IRON~\cite{iron}, which
taxonomizes per-block-type detect/recover policies---but IRON is explicitly
manual and policy-level, not a deterministic controlled experiment. Themis's
wedge is \emph{determinism makes n=1 causally sufficient}: because the fork
is a bit-exact $\varepsilon$-copy, the unflipped counterfactual is identical
modulo one bit, so every downstream divergence is \emph{provably caused} by
that bit---an exact causal blast radius rather than a failure-policy
fingerprint.

For the clock experiments, a reviewer's first instinct is that clock faking
is solved by \texttt{libfaketime}~\cite{libfaketime}. It is not: \texttt{LD\_PRELOAD}
fakes what \emph{userspace reads} via \texttt{time()}/\texttt{fstat()}; it
never touches the in-kernel timestamp-generation path that writes
\texttt{\_\_le32 i\_mtime} to disk, and it provides no exploration and no
oracle. Deterministic hypervisors such as Antithesis~\cite{antithesis}
virtualize the clock to \emph{remove} time nondeterminism for
reproducibility---the inverse posture; determinism is not exploration. No
prior filesystem tester drives a virtual clock as an explored dimension over
an unmodified in-kernel filesystem. The live, twice-reverted Linux
multigrain-ctime defect of 2023--24~\cite{multigrain}---caught in the wild
by \texttt{make}/\texttt{rsync} breakage, not by any tool---is the existence
proof that this bug class is real and current.

\label{sec:forkcommodity}
The fork substrate itself is commodity, and that is fine.
Lightweight VM fork/snapshot with copy-on-write or dirty-page reset is
standard: Potemkin~\cite{potemkin}, SnowFlock~\cite{snowflock},
Remus~\cite{remus}, Firecracker snapshots~\cite{firecracker}, Xen
\texttt{fork-vm} for fuzzing~\cite{xenfork}, Agamotto's ``lightweight VM
checkpointing as a primitive'' for a checker~\cite{agamotto}, and Nyx's
dirty-page reset~\cite{nyx}. Themis does not claim the fork. It claims (a)
that a machine-layer fork serves this specific task from beneath the
OS---where Metis, evaluating VM snapshotting within an in-OS architecture,
found it unsuitable---and (b) everything the fork \emph{enables} below and
around an unmodified filesystem. Likewise, hash-chained
tamper-evident logs are pre-empted by PeerReview~\cite{peerreview}; my
block-write chain is quiet infrastructure for future crash-epoch replay, not
a headline.

% ============================================================
\section{Threat Model, Scope, and Platform}
\label{sec:scope}

Themis tests correctness and robustness properties of an \emph{unmodified
in-kernel} filesystem: the abstract filesystem state (path, kind, mode, uid, gid,
nlink, size, content) reached by a sequence of operations; the
below-filesystem I/O behavior (what the driver writes/reads and where);
on-disk state determinism modulo nuisance fields; timestamp behavior under
an adversarial clock; and the blast radius of controlled below-filesystem
corruption. It is equally important to say what Themis is not: it is not a
crash-consistency tester (\S\ref{sec:crash}). It is not a fuzzer
of the syscall ABI for memory-safety bugs. It runs
single-node and single-vCPU.

As for the platform, Themis is built on the AnimaOS bare-metal Type-1
AMD-V/SVM hypervisor. The
host is an AMD Zen5 Ryzen 9800X3D, self-hosting from NVMe. The guest is a
stock Linux kernel with a busybox initramfs, booted single-vCPU
(\texttt{maxcpus=1}), running the stock in-kernel ext2 driver over a
virtio-blk device whose backing store is a Themis copy-on-write overlay
(\S\ref{sec:design}). Single-vCPU is deliberate: it is what makes the fork
snapshot atomic (no live AP tears the snapshot) and sidesteps cross-CPU TLB
coherence---it is also the regime a serial-console FS model checker wants.
All silicon results are from a campaign of over 40 hardware runs across four
days; each run is one physical boot of the machine with a specific build
configuration, and the raw serial/NVMe bootlogs are archived alongside this
draft.

The choice to build on a Type-1 hypervisor that \emph{owns} the machine---
rather than QEMU/KVM---is the whole point: the checker is the most
privileged software on the box, so the block device, the clock, the nested
page tables, and the VMCB are directly under its control, with no host OS
mediating. A companion QEMU ``MockFs/MockNpt'' harness proves every bit of
\emph{pure logic} without touching hardware; the hypervisor half, which
cannot run nested under QEMU (no nested SVM), is proven on silicon.

% ============================================================
\section{Design: The $\varepsilon$-Copy Fork}
\label{sec:design}

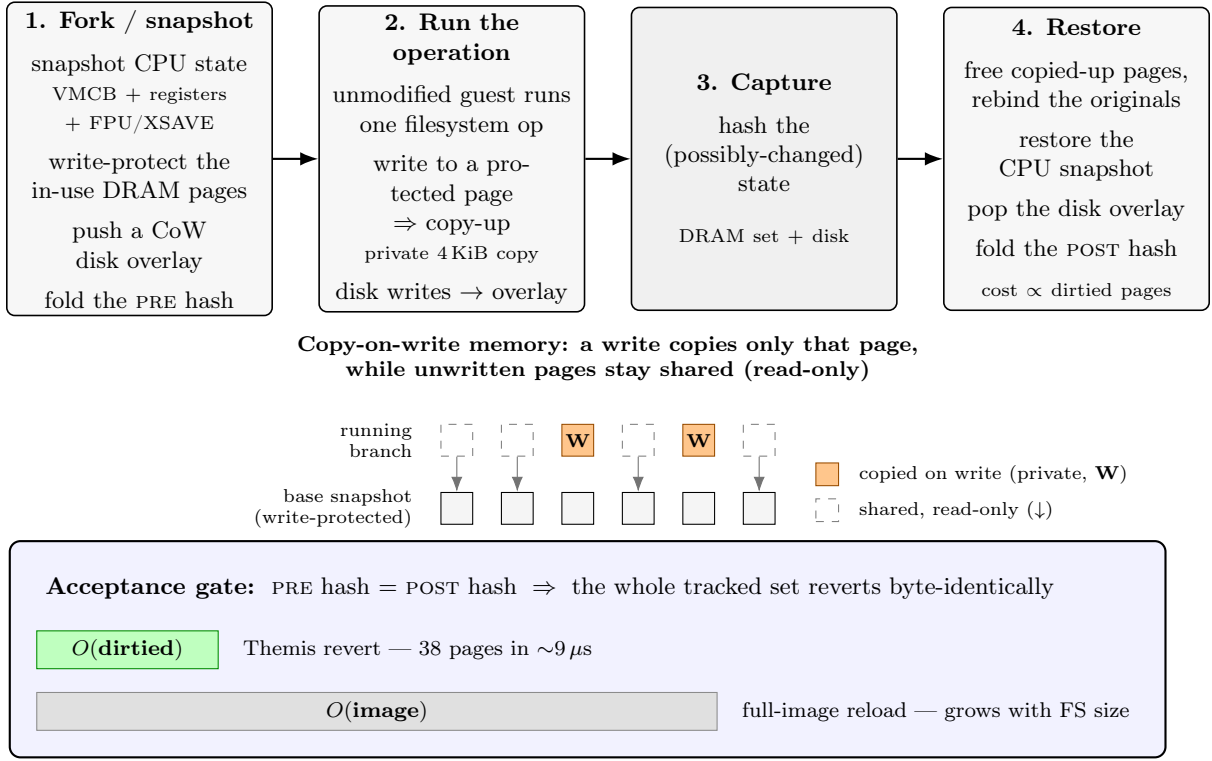
\begin{figure*}[t]\centering
\begin{tikzpicture}[
  font=\small,
  >={Latex[length=2.4mm]},
  stage/.style={draw, semithick, rounded corners=3pt, align=center,
                text width=33mm, minimum height=40mm, fill=black!3, inner sep=3pt},
  capstage/.style={stage, fill=black!6},
  cell/.style={draw, minimum size=4.2mm, inner sep=0pt, fill=black!4},
  cow/.style={cell, fill=orange!45, draw=orange!60!black},
  bar/.style={draw, minimum height=5mm, inner sep=2pt, font=\footnotesize,
              anchor=north west},
  payoff/.style={draw, thick, rounded corners=4pt, fill=blue!5},
  flow/.style={->, thick}]

  % ---- pipeline: four stages, left to right ----
  \node[stage] (fork)
    {\textbf{1. Fork / snapshot}\\[5pt]
     snapshot CPU state\\{\scriptsize VMCB $+$ registers $+$ FPU/XSAVE}\\[4pt]
     write-protect the\\ in-use DRAM pages\\[4pt]
     push a CoW disk overlay\\[4pt]
     fold the \textsc{pre} hash};
  \node[stage, right=6mm of fork] (run)
    {\textbf{2. Run the operation}\\[5pt]
     unmodified guest runs\\ one filesystem op\\[4pt]
     write to a protected page\\ $\Rightarrow$ copy-up\\{\scriptsize private 4\,KiB copy}\\[4pt]
     disk writes $\to$ overlay};
  \node[capstage, right=6mm of run] (capn)
    {\textbf{3. Capture}\\[5pt]
     hash the\\ (possibly-changed)\\ state\\[8pt]
     {\scriptsize DRAM set $+$ disk}};
  \node[stage, right=6mm of capn] (rest)
    {\textbf{4. Restore}\\[5pt]
     free copied-up pages,\\ rebind the originals\\[4pt]
     restore the CPU snapshot\\[4pt]
     pop the disk overlay\\[4pt]
     fold the \textsc{post} hash\\[4pt]
     {\scriptsize cost $\propto$ dirtied pages}};

  \draw[flow] (fork.east) -- (run.west);
  \draw[flow] (run.east)  -- (capn.west);
  \draw[flow] (capn.east) -- (rest.west);

  % ---- copy-on-write memory: shared parent pages vs copied-up private pages ----
  \coordinate (cowc) at ($(run.south)!0.5!(capn.south)$);
  \node[font=\footnotesize\bfseries, align=center, below=2mm of cowc, anchor=north] (cowtitle)
    {Copy-on-write memory: a write copies only that page,\\
     while unwritten pages stay shared (read-only)};
  % two aligned rows of six pages: child (branch) above, parent (snapshot) below
  \coordinate (chrow) at ([xshift=-20mm, yshift=-6.5mm]cowtitle.south);
  \foreach \i in {0,...,5} {
    \coordinate (cc\i) at ($(chrow)+(\i*8mm,0)$);       % child cell centre
    \coordinate (pp\i) at ($(chrow)+(\i*8mm,-9mm)$);    % parent cell centre
  }
  % parent (base) row: every page present, write-protected, shared read-only
  \foreach \i in {0,...,5}
    \node[cell] (par\i) at (pp\i) {};
  % child (branch) row: unwritten cols are hollow shares; written cols {2,4} copied-up
  \foreach \i in {0,1,3,5}
    \node[cell, fill=none, dashed, draw=black!50] (ch\i) at (cc\i) {};
  \foreach \i in {2,4}
    \node[cow] (ch\i) at (cc\i) {\scriptsize\bfseries W};
  % shared pages: no copy --- the branch reads the parent's page below it
  \foreach \i in {0,1,3,5}
    \draw[-{Latex[length=2mm]}, black!55] (ch\i.south) -- (par\i.north);
  % row labels (left of each row)
  \node[font=\scriptsize, align=right, left=2.5mm of ch0, anchor=east] {running\\branch};
  \node[font=\scriptsize, align=right, left=2.5mm of par0, anchor=east]
    {base snapshot\\(write-protected)};
  % legend (right, between the two rows)
  \coordinate (lg) at ([xshift=9mm]$(cc5)!0.5!(pp5)$);
  \node[cow, minimum size=3mm] (lgA) at (lg) {};
  \node[right=1.4mm of lgA, font=\scriptsize, anchor=west] {copied on write (private, \textbf{W})};
  \node[cell, fill=none, dashed, draw=black!50, minimum size=3mm, below=1.6mm of lgA] (lgB) {};
  \node[right=1.4mm of lgB, font=\scriptsize, anchor=west] {shared, read-only ($\downarrow$)};

  % ---- payoff: acceptance gate + O(dirtied) vs O(image) ----
  \coordinate (px) at ([yshift=-54mm]fork.west);
  \node[anchor=north west, font=\small] (gate) at ([xshift=4mm]px)
    {\textbf{Acceptance gate:}\ \ \textsc{pre} hash $=$ \textsc{post} hash
     $\;\Rightarrow\;$ the whole tracked set reverts byte-identically};
  \node[bar, fill=green!25, draw=green!55!black, minimum width=24mm] (b1)
     at ([yshift=-3mm]gate.south west) {\textbf{$O(\text{dirtied})$}};
  \node[right=2mm of b1, font=\footnotesize, anchor=west] (b1t)
    {Themis revert --- 38 pages in ${\sim}9\,\mu$s};
  \node[bar, fill=black!12, draw=black!45, minimum width=90mm] (b2)
     at ([yshift=-3mm]b1.south west) {\textbf{$O(\text{image})$}};
  \node[right=2mm of b2, font=\footnotesize, anchor=west] (b2t)
    {full-image reload --- grows with FS size};
  \begin{scope}[on background layer]
    \node[payoff, fit=(gate)(b1)(b1t)(b2)(b2t), inner sep=3.5mm] {};
  \end{scope}
\end{tikzpicture}
\caption{\textbf{One fork's lifecycle}, left to right: fork/snapshot, run one
  filesystem operation, capture, and restore. The center panel shows the
  copy-on-write memory---a write copies only that one page, while unwritten
  pages stay shared read-only. The acceptance gate is
  $\text{PRE-hash}=\text{POST-hash}$: the whole tracked set reverts
  byte-identically, so the revert is $O(\text{dirtied})$ (38 pages in
  ${\sim}9\,\mu$s) rather than a full-image reload. \S\ref{sec:design}
  details each stage.}
\label{fig:fork}
\end{figure*}

The heart of Themis is a fork whose restore cost is proportional to the
pages a branch dirtied, not the size of the filesystem image. I call it an
\textbf{$\varepsilon$-copy} fork: the child branch shares the parent's pages
until it writes one, and a restore frees only what the child privately
copied. This section describes the five pieces---the NPT write-protect +
copy-up core, the CPU/device snapshot, the copy-on-write disk overlay, the
doorbell interface, the whole-set acceptance gate---and then why they
together let every unmodified filesystem be its own reference.
Figure~\ref{fig:fork} shows one fork's lifecycle.

\label{sec:npt}
At the core of the fork is NPT write-protection plus copy-up.
On AMD-V, the guest's physical memory is mapped through nested page tables
(NPT). Themis installs guest DRAM as 4\,KiB NPT leaves on demand (the
platform does not use 2\,MiB nested PDEs---a Zen5 consumer-silicon quirk
means \texttt{PS=1} in nested PD entries is not reliably honored, so all
guest DRAM is 4\,KiB leaves; this happens to make write-protection uniform).
To fork, Themis walks the currently-installed DRAM leaves and clears the
writable bit, recording each in a tracked set (\texttt{CowForkState}).
Clearing W leaves the accessed/dirty bits untouched, so the fault that
follows is a clean permission fault, not a dirty-bit walk pathology.

A subsequent guest write to a write-protected present leaf raises a nested
page fault with error bits \texttt{P=1, R/W=1} (\texttt{err \& 0b11 ==
0b11}). Themis intercepts this at the top of the NPF handler, \emph{before}
MMIO dispatch (a tracked DRAM page is never MMIO), and performs a
\textbf{copy-up}---reading the current page contents through the host mapping,
allocating a fresh private frame, copying the 4\,KiB, remapping the leaf to the
private frame writable, recording the (gpa $\to$ private frame, parent frame)
mapping, and resuming the guest \emph{without advancing RIP} so it retries the
write against the now-private page. The parent frame is never mutated---that
immutability is what makes the fork revertible.

Two subtleties cost real time on hardware.
First, clearing the writable bit in an NPT leaf does \textbf{not} evict a
cached writable TLB entry on AMD SVM. Without a TLB flush at fork time, every
page whose translation was TLB-resident-writable at the fork point (the hot
working set---kernel stack, \texttt{current}, live page tables) is written
with \emph{no} fault, \emph{no} copy-up, and the write lands in the shared
parent frame and is never reverted. This silently corrupts the reference for
all later branches, and it is invisible to any test that only touches cold
pages. The fix is a one-shot \texttt{FLUSH\_ALL\_ASID} in the fork path.
Second, the on-demand install of a page \emph{first touched after the fork}
(page-cache growth---the commonest FS side effect) has no parent version;
restore must \textbf{unmap} it, free it, and untrack it, not merely free a
frame. The restore action set therefore carries four lists---copy-ups to
free-and-rebind, and fresh pages to unmap---validated by a MockNpt replay in
software before any hardware run. Restore reverts the NPT: free the child's
private frames, rebind the parent
leaves, unmap fresh installs, one-shot TLB flush. Because it touches only
what the branch dirtied, it is \textbf{O(dirtied)}.

\label{sec:cpusnap}
Beyond memory, a fork must capture all machine state that carries across a
branch. On AMD-V
the minimal correct set is four things: the whole 4\,KiB VMCB page (the
instruction pointer, the control and segment registers, and the segment
bases), the host-managed general-purpose registers (which, on AMD-V, live
outside the VMCB and so must be captured separately), the shadow CR2 the run
loop reloads each iteration, and the \textbf{guest XSAVE area} (the 4032-byte
block of FPU/AVX state)---the piece a naive ``copy the VMCB'' misses,
producing the classic ``fork looks fine, then floats diverge'' bug. Missing
any one of these four is a silent state leak across a branch: the memory
reverts, but a register, a segment base, or a float lane carries a value
forward from the wrong branch, so I snapshot and restore all four as a unit.
Snapshotting the \emph{whole} VMCB page is safe because the host's own save
area sits elsewhere and the fork-invariant control structures do not change
underneath it.

Once a branch performs real I/O, device state must also snapshot and revert
(gated behind a compile-time flag so a pure-compute fork pays nothing): the
whole virtio-blk device struct including \texttt{last\_avail\_idx} (the
host's consume cursor---omitting it desynchronizes the virtqueue rings
across a restore, the classic ring-desync bug), both LAPICs (pending IRR,
timer fields), the IOAPIC redirection table, and the block completion-IRQ
latches. I snapshot \emph{whole structs}, not register subsets, because the
guest-visible ring indices in DRAM and the host-shadow cursor in the device
struct must revert in lockstep. The fork is taken only at a clean VMEXIT
boundary with no in-flight virtio-blk descriptor (\texttt{used == avail})
and guest interrupts sane, so no request is mid-flight when the snapshot is
taken.

\label{sec:cow}
The disk side of the fork is copy-on-write ownership.
The virtual block device is backed by a copy-on-write structure
(\texttt{CowDisk}): an immutable base image plus a \emph{stack} of overlay
layers. \texttt{fork()} pushes a layer (O(1)); the running branch's writes
land only in the top layer; \texttt{discard()} pops it (O(1) backtrack); a
read walks the stack top-down, returning the first hit, else the base. The base image
is never mutated, so it is the permanent root of the exploration tree. This
is the disk-side $\varepsilon$-copy: a branch's disk footprint is exactly
the sectors it wrote, and reverting it is a pop. The disk overlay and the
DRAM tracked set are kept in lockstep---a fork pushes both, a restore pops
both, and the code asserts equal depth---because a filesystem write dirties
both DRAM (buffer heads, page cache) and disk (the eventual bio), and both
must revert together or the branch leaks state forward. Because the checker
owns this structure, it is also the instrument for
\S\ref{sec:observe}--\S\ref{sec:e2}: the top layer's key set \emph{is} the
below-FS write set and its LBAs; an interior read counter \emph{is}
the below-FS read amplification; a whole-image fold \emph{is} the disk
oracle for the fork's revert proof and for the byte-level state diff;
and a direct sector write \emph{is} the below-FS corruption injector.

\label{sec:doorbell}
The guest agent and the host checker rendezvous through a port-I/O doorbell:
the guest issues an \texttt{OUT} to the reserved port \texttt{0xD0}, which
VMEXITs straight to the hypervisor, and the command byte in the accompanying
register selects the action. This is the entire guest-to-host interface---no
shared memory region, no paravirtual driver, nothing the unmodified
filesystem can see---which is what keeps the guest genuinely unmodified apart
from a thin agent that rings the bell. The commands the mechanism needs are
minimal: \textbf{fork} (snapshot +
write-protect + push overlay), \textbf{restore} (revert), a \textbf{disk-mid
capture} (the guest signals its write is durable so the host can sample the
in-window disk state as a positive control), and an \textbf{exit} that tears
down to the host shell and resets fork state. Table~\ref{tab:doorbell} is
the full command map, verified byte-for-byte against the dispatcher. A
restore is a \emph{single-pass terminate}: it reverts DRAM + disk + devices
and sets RIP past the doorbell, then tears down---it does \textbf{not} loop
the guest back to re-run the op (an earlier two-pass design re-ran the FS op
against a popped overlay and corrupted the shared baseline; the revert proof
is host-side, so the guest re-run is pure liability). A fork-window watchdog
bounds VMRUN iterations so a wedged guest yields a \emph{log}, not a bricked
machine. The host-driven exploration loop that would inject operations and read back
guest fingerprints over a shared channel (needed for the at-scale throughput
pillar---scaling the backtrack cost across a large exploration) is designed
but deferred (\S\ref{sec:limitations});
a guest-scripted driver realizes the eval results in this paper without it.

\begin{table*}[t]\centering\small
\caption{The doorbell command interface. Each command is a guest
  \texttt{OUT} to a reserved port that traps to the hypervisor.}
\label{tab:doorbell}
\begin{tabularx}{\textwidth}{@{}lX@{}}
\toprule
command & what it does\\
\midrule
\texttt{0xE0} & Flip one bit in a marked file data block (the silent-corruption injector).\\
\texttt{0xE1} & Push a disk overlay and reset the read counters, establishing a per-operation baseline.\\
\texttt{0xE2} & Report the number of sectors the operation wrote below the filesystem.\\
\texttt{0xE3} & Report the sorted list of disk locations the operation wrote.\\
\texttt{0xE4} & Report the number of sectors read below the filesystem since the baseline.\\
\texttt{0xE5} & Hash the entire disk image.\\
\texttt{0xE6} & Snapshot the entire disk for a later diff.\\
\texttt{0xE7} & Report the disk sectors that differ from the snapshot.\\
\texttt{0xE8} & Report the differing byte ranges, then mask the nuisance fields and recount.\\
\texttt{0xE9} & Flip a bit of the superblock magic at LBA 2 (the fatal-corruption injector).\\
\texttt{0xEA} & Corrupt an inode-table byte.\\
\texttt{0xEB} & Corrupt a directory-block byte.\\
\texttt{0xF1} & Fork: snapshot the CPU, write-protect the working set, and push a disk overlay.\\
\texttt{0xF2} & Restore: revert memory, disk, and CPU to the fork point.\\
\texttt{0xF3} & Capture the in-window disk state as a positive control.\\
\texttt{0xFF} & Tear down fork state and return to the host shell.\\
\texttt{0x00}--\texttt{0xFE} & Progress marker, no side effect.\\
\bottomrule
\end{tabularx}
\end{table*}

\label{sec:ownref}
The payoff of owning the block device is that every filesystem becomes its
own reference, removing the need for a hand-coded one.
Metis builds RefFS because cheap snapshotting of a \emph{real} filesystem's
in-kernel state is not available to an in-OS process. Themis can: the $\varepsilon$-copy fork snapshots
and restores the whole machine---page cache, inode cache, block device---in
time proportional to the dirtied set. So the abstract state can be read
\emph{by observing the real filesystem} rather than by consulting a
hand-coded model. The checker walks the mounted filesystem
(\texttt{readdir} + \texttt{stat} + content read) and folds each entry into
a Merkle-style \textbf{fingerprint}---a sorted \texttt{path} $\to$ leaf-hash
map whose fold is a single deterministic root---over exactly Metis's policy
fields (path, kind, mode, uid, gid, nlink, size, content), \emph{excluding}
time by default so that the substrate delta is isolated from the clock
delta (a config folds mtime in for the clock experiments). Two questions a checker lives on
both reduce to this fingerprint: ``have we seen this state?'' is a lookup in
a visited set keyed by root hash (O(states), not O(image)); ``do two
implementations agree?'' is a root comparison, and the first sorted path
where they differ \emph{is} the localized differential bug.

\begin{figure*}[t]\centering
\begin{tikzpicture}[
  font=\small, node distance=8mm and 24mm,
  box/.style={draw, rounded corners=2pt, align=left, text width=56mm, fill=black!3, inner sep=4pt},
  ref/.style={draw, dashed, rounded corners=2pt, align=left, text width=56mm, inner sep=4pt},
  >={Latex[length=2mm]}]

  \node[box] (mk)
    {\textbf{mke2fs a real ext2} in the guest,\\ then a deterministic,
     replayable operation sequence};
  \node[box, below=of mk] (obs)
    {\textbf{Below-FS observers} (own the device):\\ write amplification,
     write localization, read amplification\\ (below-the-device measurements)};
  \node[box, below=of obs] (snap)
    {\textbf{Whole-disk snapshot, diff,}\\ \textbf{and nuisance normalization}};

  \node[box, right=of mk] (fp)
    {\textbf{Read-oracle fingerprint}:\\ walk names $\to$ stat inode $\to$
     hash content;\\ leaf $=$ path $|$ kind $|$ mode $|$ uid $|$ gid $|$
     nlink $|$ size $|$ content\\ (mtime folded only in the time mode);\\
     Merkle-fold to a root; dedup against the visited set};
  \node[ref, below=of fp] (diff)
    {\textbf{Every filesystem is its own reference}:\\ drive two instances
     through the same sequence;\\ the diff localizes the first divergence\\
     \emph{(no hand-coded reference)}};
  \node[box, below=of diff] (dfs)
    {\textbf{Exploration DFS loop}:\\ fingerprint $\to$ dedup $\to$ fork
     $\to$ apply op $\to$ restore};

  \node[draw, rounded corners=2pt, fill=black!8, align=center,
    text width=110mm, anchor=north]
    (v) at ([yshift=-14mm]$(snap.south)!0.5!(dfs.south)$)
    {\textbf{Verdict}: divergence path $\cdot$ below-FS footprint tables
     $\cdot$ deterministic-modulo-nuisance normalization $\cdot$ corruption
     blast radius};

  \draw[->] (mk.south) -- (obs.north); \draw[->] (obs.south) -- (snap.north);
  \draw[->] (mk.east) -- (fp.west);  \draw[->] (fp.south) -- (diff.north); \draw[->] (diff.south) -- (dfs.north);
  \draw[->] (snap.south) -- (snap.south |- v.north);
  \draw[->] (dfs.south)  -- (dfs.south |- v.north);
\end{tikzpicture}
\caption{\textbf{The data-collection loop.} A run creates a real ext2 with
  \texttt{mke2fs} and drives a deterministic, replayable operation sequence
  through two parallel planes---the below-FS observers and the whole-disk
  snapshot/diff/normalize path---while the read-oracle fingerprint dedups
  states and the exploration loop brackets every operation with
  fork/apply/restore. Driving two unmodified instances through the same
  sequence localizes the first divergent path, removing the need for a
  hand-coded reference (\S\ref{sec:ownref}). All paths converge on a verdict.}
\label{fig:pipeline}
\end{figure*}
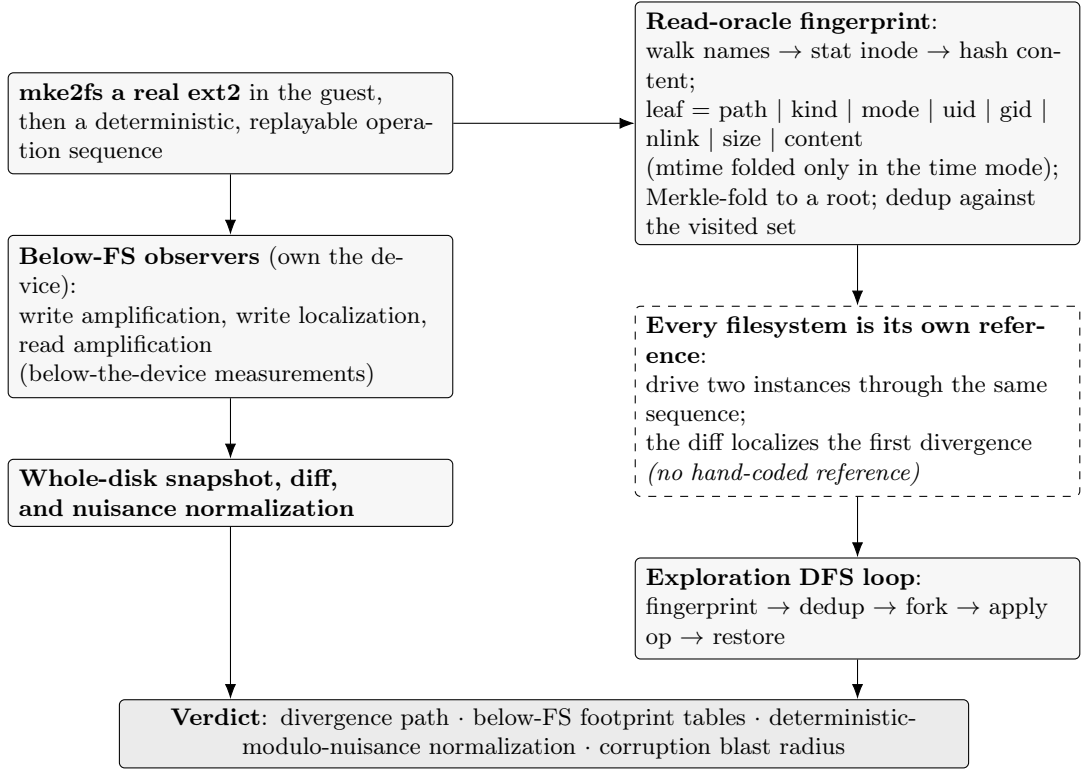

That second question is what removes the need for a hand-coded reference: drive two \emph{unmodified}
filesystem instances (two implementations, or one instance versus its
remounted self) through the same deterministic operation sequence; a root
divergence at any step is a bug in one relative to the other, found with
\textbf{no hand-built oracle}. The exploration driver is a bounded DFS: at
each state, apply \texttt{next\_command} from a deterministic policy (same
seed $\Rightarrow$ same exploration $\Rightarrow$ replayable), fork before
each op, recurse if the op changed the abstract state, restore after---every
op is fork/restore-bracketed, and the tree unwinds fully
(Figure~\ref{fig:pipeline}). The DFS, the visited-set dedup, the
differential comparison, and the fingerprint are all pure logic, proven
in software under QEMU against an in-memory \texttt{MockFs}/\texttt{RealisticFs};
the real checker swaps the mock for the hypervisor fork plus a guest
read-oracle whose content digest is a \texttt{blake3} over the file's
\emph{real bytes} (the model carries the actual content), so a
size-preserving content corruption changes the fingerprint and is
caught---an earlier draft digested only \texttt{path}~$\|$~\texttt{size}
and was blind to it, and that gap is now closed.

% ============================================================
\section{Machine-Owned Observation}
\label{sec:observe}

An in-OS checker reads syscall arguments and returns. Themis owns the block
device, so for every filesystem operation it measures the
\emph{below-filesystem} consequences directly. These are not incidental
metrics; they are a data class the beneath-the-device vantage makes
available, and they let Themis reverse-engineer the on-disk layout and
compare whole-disk states at byte granularity.

\label{sec:profiles}
The first data class is the below-FS I/O profile of each operation---its
write amplification, its write localization, and its read amplification.
Consider write amplification first.
For each operation Themis counts the sectors written beneath the filesystem
(the top overlay layer's key set). Table~\ref{tab:n1} shows the profile for
real ext2. A 1-byte write costs 16 sectors (8192$\times$ amplification); the
amplification falls toward $\sim$1$\times$ as the write grows past the block
size; and \emph{metadata} operations that a syscall-return checker sees as
``returned 0'' have a hidden, non-zero below-FS cost (create/mkdir = 14
sectors, chmod/chown = 2, hardlink/rename/unlink = 6). What this profile
means is that the syscall return code is a lossy summary of what the
filesystem actually did: a metadata operation the guest reports as a bare
success in fact commits a fixed, characteristic set of sectors below the
driver, and the amplification curve---8192$\times$ for a sub-block write,
decaying to unity once the write spans whole blocks---directly traces the
block-size quantum and the read-modify-write it forces. Each operation
therefore has a below-FS write \emph{signature}, and that signature is a
first-class observable Themis can key on, deduplicate against, and diff---a
dimension of behavior that the syscall return value does not by itself
express.

\begin{table}[t]\centering\small
\caption{Below-FS write amplification for real ext2.}
\label{tab:n1}
\resizebox{\columnwidth}{!}{%
\begin{tabular}{@{}lrrr@{}}
\toprule
op & bytes & sectors written & amplification\\
\midrule
write 1\,B & 1 & 16 & 8192$\times$\\
write 512\,B & 512 & 16 & 16$\times$\\
write 4\,KB & 4096 & 20 & 2.5$\times$\\
write 64\,KB & 65536 & $\sim$130 & $\sim$1.02$\times$\\
write 1\,MB & 1048576 & 2070 & $\sim$1.01$\times$\\
create / mkdir & 0 & 14 & ---\\
chmod / chown & 0 & 2 & ---\\
hardlink / rename / unlink & 0 & 6 & ---\\
\bottomrule
\end{tabular}}
\end{table}

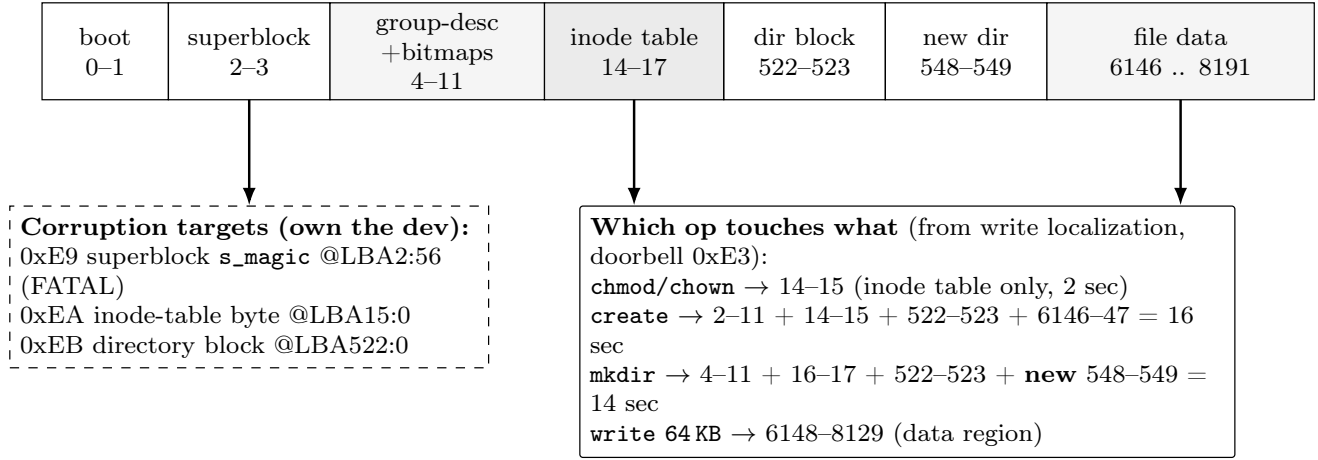
\begin{figure*}[t]\centering
\resizebox{\textwidth}{!}{%
\begin{tikzpicture}[
  font=\footnotesize, node distance=0mm,
  reg/.style={draw, minimum height=11mm, align=center, outer sep=0},
  op/.style={-{Latex[length=2mm]}, thick},
  cor/.style={draw, dashed, align=left, text width=52mm},
  >={Latex[length=2mm]}]

  \node[reg, text width=12mm] (boot) {boot\\ 0--1};
  \node[reg, text width=16mm, right=of boot] (sb) {superblock\\ 2--3};
  \node[reg, text width=22mm, right=of sb, fill=black!4] (gd) {group-desc\\+bitmaps\\ 4--11};
  \node[reg, text width=18mm, right=of gd, fill=black!8] (it) {inode table\\ 14--17};
  \node[reg, text width=16mm, right=of it] (dir) {dir block\\ 522--523};
  \node[reg, text width=16mm, right=of dir] (nd) {new dir\\ 548--549};
  \node[reg, text width=28mm, right=of nd, fill=black!4] (data) {file data\\ 6146 .. 8191};

  \coordinate (gtop) at ([yshift=-12mm]$(boot.south west)!0.5!(data.south east)$);
  \coordinate (opsc) at ($(it)!0.5!(data)$);

  \node[draw, rounded corners=1pt, align=left, text width=72mm, anchor=north] (ops)
    at (opsc |- gtop)
    {\textbf{Which op touches what} (from write localization, doorbell 0xE3):\\
     \texttt{chmod/chown} $\to$ 14--15 (inode table only, 2 sec)\\
     \texttt{create} $\to$ 2--11 + 14--15 + 522--523 + 6146--47 = 16 sec\\
     \texttt{mkdir} $\to$ 4--11 + 16--17 + 522--523 + \textbf{new} 548--549 = 14 sec\\
     \texttt{write 64\,KB} $\to$ 6148--8129 (data region)};
  \node[cor, anchor=north] (cor) at (sb |- gtop)
    {\textbf{Corruption targets (own the dev):}\\
     0xE9 superblock \texttt{s\_magic} @LBA2:56 (FATAL)\\
     0xEA inode-table byte @LBA15:0\\
     0xEB directory block @LBA522:0};

  \draw[op] (sb.south)   -- (cor.north);
  \draw[op] (it.south)   -- (it.south   |- ops.north);
  \draw[op] (data.south) -- (data.south |- ops.north);
\end{tikzpicture}}
\caption{\textbf{The ext2 layout, reverse-engineered without reading a line
  of filesystem code.} Every region boundary is derived from watching which
  LBAs each operation dirties (Table~\ref{tab:n2}); the corruption doorbells
  (\S\ref{sec:e2}) target exactly these structures. Owning the block device
  recovers the on-disk format from below---a view available from beneath the
  device, complementary to the syscall interface an in-OS checker reads.}
\label{fig:ext2}
\end{figure*}

Write localization reverse-engineers the ext2 layout from beneath.
Because Themis sees \emph{which} LBAs each op writes, it derives the on-disk
layout without reading a line of filesystem source (Table~\ref{tab:n2}). From
these I read off, over the 8192-sector 4\,MiB image: LBA 2--3 = superblock,
4--11 = group descriptors + bitmaps, 14--17 = inode table, 522--523 /
548--549 = directory blocks, 6146+ = file data. A chmod touches \emph{only}
the inode table (LBAs 14--15); a create touches the superblock, bitmaps,
inode table, a directory block, and a data block; a mkdir additionally
allocates a new directory block (548--549). What makes this more than a
curiosity is that the entire on-disk format---region boundaries, which
structure each operation mutates, and where---falls out purely from watching
below-FS writes, and the same map then tells the corruption experiments
(\S\ref{sec:e2}) exactly which LBA and byte offset to target.
Figure~\ref{fig:ext2} draws the full reverse-engineered map.

\begin{table*}[t]\centering\small
\caption{Below-FS write localization, and the reverse-engineered ext2
  layout.}
\label{tab:n2}
\begin{tabular}{@{}lrll@{}}
\toprule
op & sectors & LBAs touched & ext2 structure\\
\midrule
chmod / chown & 2 & 14--15 & inode table only\\
create & 16 & 2--11, 14--15, 522--523, 6146--47 & sb $\to$ bitmaps $\to$ inode $\to$ dir $\to$ data\\
mkdir & 14 & 4--11, 16--17, 522--523, \textbf{548--549} & + new dir block\\
write 64\,KB & 142 & \ldots 6148--8129 & data region\\
\bottomrule
\end{tabular}
\end{table*}

Read amplification is the mirror of the write set: cold reads beneath the
filesystem, minus a 70-sector mount baseline (Table~\ref{tab:n2b}). A \texttt{stat} reads 2
sectors (the inode); a small \texttt{cat} reads 4 (inode + data); a 64\,KB
\texttt{cat} reads 132 (inode + 128 data + indirect block); a full
\texttt{find} over a tree reads 28. An honesty note: \texttt{drop\_caches}
was a no-op on this guest, so cold reads were forced by \texttt{umount} +
\texttt{remount} rather than by dropping caches---a coverage gap
(\S\ref{sec:limitations}) but not an inaccuracy in the reported counts.

\begin{table}[t]\centering\small
\caption{Below-FS read amplification (marginal cold reads over a
  70-sector mount baseline).}
\label{tab:n2b}
\begin{tabular}{@{}lr@{}}
\toprule
op & marginal cold reads\\
\midrule
stat & 2 (inode)\\
cat small & 4 (inode + data)\\
cat 64\,KB & 132 (inode + 128 data + indirect)\\
ls dir & 4\\
find tree & 28 (full walk)\\
chmod & 2\\
\bottomrule
\end{tabular}
\end{table}

The three profiles compose into a \textbf{3-D fingerprint per operation}---
reads $\oplus$ writes $\oplus$ location. For example,
\texttt{chmod} = read 2 / write 2 / at LBA 14--15. This is a below-FS
behavioral signature that the beneath-the-device vantage provides.

\label{sec:transition}
The filesystem can also serve as its own transition model.
Owning the block device lets Themis \emph{derive} the abstract-state
transition model by observing the real filesystem rather than hand-coding it
in a reference. Table~\ref{tab:f}, the operation--field footprint, records
for each operation exactly which of the eight abstract-state fields it
changes, measured against the real driver rather than posited. It reads as a
compact behavioral contract: chmod, chown, setuid, rename, and hardlink all
touch \texttt{ctime} (a metadata change always bumps the change time), while
only the size-changing operations (append, truncate) also touch
\texttt{mtime}, and only \texttt{touch} moves \texttt{atime} and
\texttt{mtime} together. Two entries carry a value beyond the check mark:
\texttt{setuid} sets \texttt{mode}\,=\,4755, and \texttt{hardlink}'s
\texttt{nlink} changes from 1 to 2---transitions a hand-coded reference must
encode explicitly, whereas Themis observes the real filesystem make them.
Table~\ref{tab:g}, the timestamp-semantics matrix, records the measured
timestamp semantics---and from it Themis reads, directly from the running
mount, a real mount property: this mount is \textbf{strictatime} (read
updates atime), not relatime/noatime. The point of
both tables is that the transition model is \emph{observed}, so it is correct
by construction for whatever driver, version, and mount options are actually
running, and it needs no maintenance as the filesystem evolves---which is the
precise burden RefFS carries in Metis.

\begin{table*}[t]\centering\small
\caption{The operation--field footprint (measured on real ext2): op
  $\times$ 8-field metadata footprint, where a check mark means the field
  changed.}
\label{tab:f}
\begin{tabular}{@{}lcccccccc@{}}
\toprule
op & mode & uid & gid & size & nlink & atime & mtime & ctime\\
\midrule
chmod    &          &          &          &          &          &          &          & \checkmark\\
chown    &          & \checkmark & \checkmark &        &          &          &          & \checkmark\\
setuid   & \checkmark &         &          &          &          &          &          & \checkmark\\
append   &          &          &          & \checkmark &        &          & \checkmark & \checkmark\\
truncate &          &          &          & \checkmark &        &          & \checkmark & \checkmark\\
touch    &          &          &          &          &          & \checkmark & \checkmark & \checkmark\\
rename   &          &          &          &          &          &          &          & \checkmark\\
hardlink &          &          &          &          & \checkmark &        &          & \checkmark\\
\bottomrule
\end{tabular}
\end{table*}

\begin{table}[t]\centering\small
\caption{The timestamp-semantics matrix (canonical POSIX, measured on real
  ext2).}
\label{tab:g}
\begin{tabular}{@{}lccc@{}}
\toprule
op & atime & mtime & ctime\\
\midrule
read  & \checkmark & --- & ---\\
write & --- & \checkmark & \checkmark\\
chmod & --- & --- & \checkmark\\
stat  & --- & --- & ---\\
\bottomrule
\end{tabular}
\end{table}

\label{sec:statespace}
A state-space battery of measured properties (Table~\ref{tab:statespace})
confirms the checker's abstract-state machinery on a real filesystem:
directory scaling is linear ($\sim$1\,KB/file); hardlink refcounts follow
the exact trajectory create=1 $\to$ link$\times$3 = 2,3,4 $\to$ unlink
$\times$2 = 3,2; limits hold (255-char names, dotfiles, depth-5); a 6-op
sequence run three times from a fresh mkfs yields an \emph{identical}
fingerprint \texttt{b4bb1041b38d} (the checker is reproducible); commutativity
is non-trivial (chmod;chown, chmod;write, trunc;chmod all commute, but
\textbf{write;truncate does not}); idempotency splits state from return code
(\textbf{mkdir is state-idempotent but its second return code is an error});
six error paths return the correct errno; and symlink semantics hold.

\begin{table*}[t]\centering\small
\caption{The state-space battery: measured state-space structure (summary).}
\label{tab:statespace}
\begin{tabular}{@{}lp{11cm}@{}}
\toprule
property & result\\
\midrule
directory scaling & $\sim$1\,KB/file, linear (10/100/500 files $\to$ 11/103/510\,KB on disk)\\
hardlink refcount trajectory & create=1 $\to$ link$\times$3=2,3,4 $\to$ unlink$\times$2=3,2 (exact)\\
limits & 255-char name OK, dotfile OK, depth-5 OK\\
determinism & same 6-op seq $\times$3 (fresh mkfs) $\to$ fingerprint \texttt{b4bb1041b38d} identical\\
commutativity & chmod;chown, chmod;write, trunc;chmod = yes; \textbf{write;truncate = no}\\
idempotency & chmod/truncate/chown = yes; \textbf{mkdir state-idempotent but rc2 = err}\\
error paths & 6, correct errno (rmdir-nonempty, rmdir-a-file, unlink/rename/link-missing, mkdir-existing)\\
symlinks & to-file deref yes, dangling no, to-dir is-dir yes, self-loop ELOOP (no)\\
\bottomrule
\end{tabular}
\end{table*}

\label{sec:bytediff}
The sharpest expression of owning the block device is a \textbf{byte-level
whole-disk state diff}. Themis folds the entire overlay-resolved disk into a
snapshot, runs a filesystem operation, and reports the exact LBAs---and
within them the exact byte ranges---that changed. Between two logically
identical runs this is the raw material for a disk-level determinism
check---one phrased in on-disk bytes rather than syscall returns. The
subtlety, which I get exactly right and report honestly
(\S\ref{sec:e1}), is \emph{nuisance}. Two logically identical
\texttt{mkfs + op} runs, even with a pinned clock and a fixed UUID request,
differ in a handful of sectors---not only in timestamps but in the
filesystem's \textbf{random-identity fields}: the superblock
\texttt{s\_uuid} and htree \texttt{s\_hash\_seed}, and each inode's
\texttt{i\_generation} (an NFS version counter seeded randomly). A model
checker must canonicalize these out of the abstract state---exactly as it
excludes atime. Themis does this at byte granularity: it masks the known
ext2 nuisance offsets (superblock
\texttt{s\_mtime}/\texttt{s\_wtime}/\texttt{s\_lastcheck}/\texttt{s\_uuid}/
\texttt{s\_hash\_seed}; per-inode
\texttt{i\_atime}/\texttt{i\_ctime}/\texttt{i\_mtime}/\texttt{i\_dtime}/
\texttt{i\_generation}, plus the extra timestamp fields for 256-byte inodes)
in both images before comparing, and counts the sectors that \emph{still}
differ. This ``normalized diff'' is the disk-level analogue of the
fingerprint's abstract-state canonicalization, and it is what makes the
clock-chapter result (\S\ref{sec:e1}) a clean 3 $\to$ 0.

% ============================================================
\section{The Clock Chapter}
\label{sec:e1}

Metis sets time aside from its abstract state, treating it as noise, and
notes that atime bugs fall outside its scope---a reasonable choice for a
checker that reads syscall returns rather than the on-disk timestamp path.
Themis owns the clock, so time is a dimension it \emph{explores}. This is the
paper's principal result because it is un-preempted ground: no prior filesystem
tester drives a virtual clock over an unmodified in-kernel filesystem, and
the bug class is live (the 2023--24 multigrain-ctime defect,
\S\ref{sec:corruptfake}).

I should be precise about what ``owning the clock'' means here in mechanism,
because it is easy to over-read. Themis does not run a bit-exact virtual
clock, and it does not intercept the guest's timestamp counter: the TSC runs
free. I drive time instead by having the guest agent call
\texttt{settimeofday} to advance \texttt{CLOCK\_REALTIME} in the guest
timekeeper to each adversarial value, then read back what the unmodified
filesystem chooses to persist to disk. The hypervisor is the ultimate time
source in that it presents the platform timers the guest calibrates against,
but the exploration lever is the ordinary POSIX wall clock, not the raw
RDTSC---freezing which would be the wrong instrument (\S\ref{sec:threetozero}).
What owning the machine adds is not a faked clock; it is the vantage
\emph{beneath} the driver to read exactly what the filesystem wrote in
response to that clock, at byte and, later in this section, nanosecond
granularity---something no in-guest test can see.

\label{sec:y2038}
The principal silicon result is a real Y2038 defect, witnessed on real ext2:
a guest mounts a real, unmodified in-kernel
ext2, sets its wall clock past 2038 (\texttt{date -s @2147483648}---
\texttt{settimeofday} into the guest timekeeper; no hypervisor clock
trickery), \texttt{touch}es a file, and reads back the on-disk
\texttt{\_\_le32 i\_mtime} via \texttt{stat}. Real ext2 \textbf{clamps} the
post-2038 timestamp to \texttt{0x7FFFFFFF} (\texttt{i32::MAX} =
2038-01-19 03:14:07): it cannot represent 2038-01-19 03:14:08 or later. The
Linux ext2 driver \emph{itself} declares this limit at mount (``supports
timestamps until 2038-01-19 (0x7fffffff)'')---irrefutable ground truth---and
the clamp \textbf{persists across a remount}, i.e. it is a real on-disk
value, not a page-cache artifact. The same clamp fires for a year-2106
timestamp. Table~\ref{tab:clocksweep} is the differential clock sweep, and it
pins the \emph{causal variable}: the clamp is a property of the
\textbf{128-byte inode}, not of ext2 versus ext4. A 128-byte inode has no
\texttt{i\_[cma]time\_extra} epoch bits, so it stores a bare signed
\texttt{\_\_le32} second and wraps at $2^{31}$ (2038); a 256-byte inode
carries the extra epoch bits and represents past it (to the $\sim$2446
ceiling). A host-side byte-parse cross-check---re-\texttt{mkfs}ing the same
formats and reading the on-disk inode bytes directly---isolates this cleanly:
ext4 formatted \texttt{-I~128} clamps at 2038 \emph{identically} to ext2, and
ext2 formatted \texttt{-I~256} stores post-2038 \emph{correctly}. So the
filesystem type is not the variable; the inode size is. The silicon
differential drives a 128-byte-inode lane (clamps) against a 256-byte-inode
lane (represents). A \emph{time-included} fingerprint (Themis) catches and
localizes the divergence at the affected file; a \emph{time-excluded}
(Metis-parity) fingerprint hashes the correct and clamped states identically
and so does not register it.

\begin{table}[t]\centering\small
\caption{Clock sweep: the Y2038 clamp tracks \emph{inode size}, not
  filesystem type. The silicon differential drives ext2 (128-byte inode)
  against ext4 (256-byte inode); a host-side byte-parse cross-check adds the
  diagonal (ext4 \texttt{-I~128} clamps, ext2 \texttt{-I~256} represents),
  isolating the 128-byte inode's missing \texttt{i\_[cma]time\_extra} epoch
  bits---not the filesystem type---as the causal variable.}
\label{tab:clocksweep}
\resizebox{\columnwidth}{!}{%
\begin{tabular}{@{}llll@{}}
\toprule
filesystem & inode size & post-2038 timestamp & provenance\\
\midrule
ext2 & 128-byte & clamps to \texttt{0x7FFFFFFF} (2038) & silicon\\
ext4 & 256-byte & represents (to $\sim$2446) & silicon\\
ext4 & 128-byte & clamps to \texttt{0x7FFFFFFF} (2038) & host parse\\
ext2 & 256-byte & represents (to $\sim$2446) & host parse\\
\bottomrule
\end{tabular}}
\end{table}

\begin{table}[t]\centering\small
\caption{Y2038: timestamp granularity and the persisted clamp.}
\label{tab:y2038}
\begin{tabular}{@{}p{3cm}p{4.7cm}@{}}
\toprule
probe & result\\
\midrule
granularity (4 rapid \texttt{touch}es) & 1 distinct stored time (coarse, 1\,s)\\
post-2038 store & \texttt{0x7FFFFFFF} clamp; kernel declares ``until 2038-01-19''\\
clamp across remount & persisted (real on-disk value)\\
\bottomrule
\end{tabular}
\end{table}

The coarse-granularity probe (Table~\ref{tab:y2038}) is itself a finding:
four rapid touches produce \textbf{one} distinct stored timestamp, and in
the multi-state exploration (\S\ref{sec:eval}) two rapid touches received the
\emph{identical} nanosecond-precision timestamp. On real silicon the kernel
timestamp is coarse, so two distinct operations are temporally
\emph{indistinguishable}---the multigrain-timestamp class ground truth. A
checker that relies on time to order operations is fooled; a time-included
fingerprint sees them as time-identical; a time-excluded (Metis-parity)
fingerprint does not track time at all.

\begin{figure}[t]\centering
\begin{tikzpicture}[
  font=\footnotesize, node distance=6mm,
  b/.style={draw, rounded corners=1pt, align=left, text width=72mm},
  >={Latex[length=2mm]}]
  \node[b] (raw) {\textbf{0xE7 raw diff:} 3 sectors differ --- LBA 2 (superblock), LBA 15,16 (inode table)};
  \node[b, below=of raw] (byte)
    {\textbf{0xE8 byte-diff localizes the bytes:}\\
     superblock: \texttt{s\_mtime}@44 \texttt{s\_wtime}@48 \texttt{s\_lastcheck}@64,\\
     \texttt{s\_uuid}@104(16), \texttt{s\_hash\_seed}@236(16) \emph{(random identity)}\\
     inode: \texttt{i\_atime}@8 \texttt{i\_ctime}@12 \texttt{i\_mtime}@16 \texttt{i\_dtime}@20,\\
     \texttt{i\_generation}@100(4), *\_extra@132 (256B inode)};
  \node[b, below=of byte, fill=black!6] (norm)
    {\textbf{0xE8 normalize} (\texttt{mask\_nuisance\_fields}):\\
     sectors still differing $= \mathbf{0}$ $\Rightarrow$ deterministic \emph{modulo nuisance}\\
     (3$\to$0 by canonicalization, NOT by freezing the clock)};
  \draw[->] (raw.south) -- (byte.north);
  \draw[->] (byte.south) -- (norm.north);
\end{tikzpicture}
\caption{\textbf{The byte-diff and normalize step (the 3$\to$0 result).}
  Two logically-identical runs differ in three raw sectors; localizing the
  differing bytes shows the residue lands on the filesystem's random-identity
  fields (\texttt{s\_uuid}, \texttt{s\_hash\_seed}, \texttt{i\_generation}),
  not only on timestamps. Masking those nuisance fields drops the count to
  \textbf{0}---deterministic \emph{modulo nuisance}, by canonicalization, not
  by freezing the clock (\S\ref{sec:threetozero}).}
\label{fig:normalize}
\end{figure}
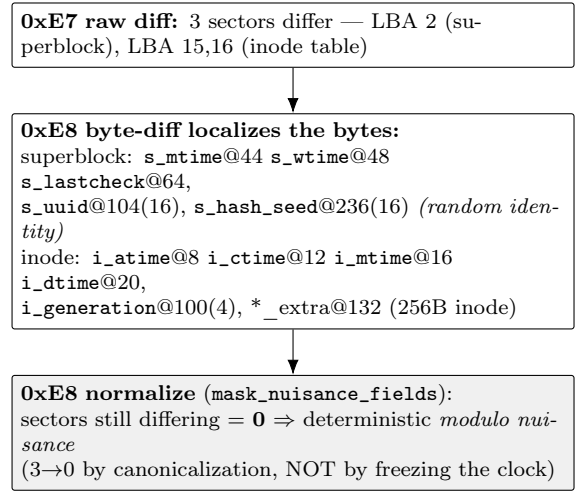

\label{sec:threetozero}
The determinism result is a 3 $\to$ 0 byte-diff, with an honest correction.
Two logically identical runs (pinned clock,
requested fixed UUID) differ \textbf{raw} in three sectors---the superblock
plus inode-table sectors. Themis's byte-diff localizes the differing bytes
and---this is the honest correction I make prominently---they are
\textbf{not (only) timestamps}. The residue is the filesystem's
random-identity fields: \texttt{s\_uuid} (the \texttt{-U} flag did not fully
pin it), \texttt{s\_hash\_seed}, and \texttt{i\_generation}. An earlier
framing of this result as ``the only difference is time'' was \emph{wrong},
and the byte-diff is what disproved it. Normalizing those nuisance fields
(\S\ref{sec:bytediff}) drives the still-differing sector count to
\textbf{zero}: the disk is deterministic \emph{modulo nuisance}
(Figure~\ref{fig:normalize}). This is exactly what a model checker does---
canonicalize nuisance out of the abstract state---performed at byte
granularity on a real disk, at the below-device level the machine owner
observes. I stress that the 3 $\to$ 0 was achieved by \textbf{canonicalization,
not by freezing the clock}.\footnote{These raw and normalized sector counts
come from a whole-disk byte-diff of the two runs' dumped disk images (the
\texttt{0xE7}/\texttt{0xE8} observation path, \S\ref{sec:bytediff}), recorded
in the evaluation dataset---an offline comparison of on-silicon runs, as
distinct from the timestamp-clamp witness (\S\ref{sec:y2038}), which prints
directly to the guest serial log.} Two adversarial design reviews unanimously rejected a
clock-freeze feature: ext2 timestamps come from
\texttt{CLOCK\_REALTIME}/\texttt{settimeofday}, not raw RDTSC (freezing the
TSC is the wrong lever); Linux's clocksource watchdog defeats a frozen TSC
within a second (operation windows are multi-second); and reaching diff=0 by
freezing would be circular. Freezing belongs instead to the
deterministic-replay engine (\S\ref{sec:replay}), where a masked-timer
freeze makes ext2 timestamps bit-exact at no additional cost. The result stands
because it is a \emph{canonicalization}, the same move the abstract state
makes for atime.

A closing extension of the clock chapter turns the same vantage on
\emph{sub-second} time, and it makes vivid both what owning the machine buys
and---just as honestly---what it does not. Because Themis owns the block
device, it can read the nanosecond ctime of a file straight from the raw
on-disk ext4 inode: a granularity available from beneath the device, whereas
the second-resolution \texttt{stat} a stock guest's userspace offers is all
that is visible from inside. I parse the ext4 superblock and group descriptor to
locate the inode, then read \texttt{i\_ctime} together with the
\texttt{i\_ctime\_extra} field whose upper bits carry the sub-second
nanoseconds. On a modern Linux guest (kernel~6.17) I made eight rapid appends
to a single file, querying and flushing between each, and read the on-disk
nanosecond ctime after every modification.

The two vantages tell opposite stories (Table~\ref{tab:multigrain}). From
inside the guest, \texttt{stat} reported a flat, unchanging whole second
throughout: the entire run fell within one wall-clock second, so at second
resolution the file never appears to change time. Read from below, the
on-disk nanosecond ctime advanced strictly and monotonically on every single
modification---from \texttt{.013921066} at creation, through
\texttt{.018000001}, \texttt{.022000002}, \texttt{.027000002},
\texttt{.031000003}, \texttt{.036000003}, \texttt{.044000004},
\texttt{.048000004}, to \texttt{.053000005} at the last append. The low-order
digits even expose the kernel's multigrain floor counter---the mechanism that
guarantees each observed timestamp is distinct within a single coarse
tick~\cite{multigrain}. This is exactly the sub-second time behavior that is
not visible from inside the guest and lies outside Metis's abstract state,
which sets time aside: a class of observation, and a class of data, that the
machine-owner vantage makes available.

\begin{table}[t]\centering\small
\caption{Nanosecond ctime read from the raw on-disk ext4 inode across eight
  rapid appends, beside the flat whole second the guest's \texttt{stat}
  reports (\S\ref{sec:e1}).}
\label{tab:multigrain}
\begin{tabular}{@{}lll@{}}
\toprule
modification & on-disk ctime (ns) & guest \texttt{stat}\\
\midrule
creation & \texttt{.013921066} & \multirow{9}{*}{\shortstack[l]{one\\second\\(flat)}}\\
append 1 & \texttt{.018000001} & \\
append 2 & \texttt{.022000002} & \\
append 3 & \texttt{.027000002} & \\
append 4 & \texttt{.031000003} & \\
append 5 & \texttt{.036000003} & \\
append 6 & \texttt{.044000004} & \\
append 7 & \texttt{.048000004} & \\
append 8 & \texttt{.053000005} & \\
\bottomrule
\end{tabular}
\end{table}

I am scrupulous about what this is and is not. It is a
capability-and-data demonstration, not a newly discovered bug. On this kernel
the multigrain implementation is \emph{correct}: every rapid modification
advanced the nanosecond ctime, and I observed no missed-change anomaly,
because the earlier multigrain-ctime defects (\S\ref{sec:corruptfake}) are
fixed in kernel~6.17. What the experiment establishes is therefore the
observation vantage and a data class the machine owner alone can produce---not
the reproduction of a previously-unknown defect. It reinforces, rather than
closes, the honest note in the limitations (\S\ref{sec:limitations}): a
previously-unknown reproduced bug remains the single strongest sentence the
evaluation still lacks, and the search---across other filesystems and
kernels---continues.

% ============================================================
\section{Causal Below-Filesystem Corruption}
\label{sec:e2}

Themis owns the block device, so it can flip a single bit in a chosen
on-disk structure---a controlled fault injection performed from beneath the
device---and the guest's re-mount/re-read reveals the blast radius. Because
the fork is a bit-exact $\varepsilon$-copy, the unflipped counterfactual is
identical modulo the one bit, so \textbf{every downstream difference is
provably caused by that bit}: determinism makes a single run a complete
causal experiment (n=1 is sufficient), in contrast to IRON's manual,
policy-level failure fingerprints.

I demonstrate a blast-radius \emph{spectrum} (Table~\ref{tab:e2}). At the
silent end, flipping one bit in a file's data block: the file reads back
wrong (a byte changes, e.g. \texttt{-} $\to$ \texttt{,}), the filesystem
reports \textbf{zero} error, and other files are intact. ext2 has no data
checksums, so the read returns success with wrong bytes---a corruption for
which a checker reading syscall returns has no signal to key on. At the fatal end, flipping one
bit of the superblock magic (LBA 2, byte 56): the whole filesystem becomes
unmountable---maximum blast radius from a single bit. The spectrum is the
result: the \emph{same} one-bit perturbation ranges from completely
undetectable to total loss of the volume purely as a function of which
on-disk structure it lands in. Because the fork makes the unflipped
counterfactual bit-exact, each blast radius is \emph{measured} rather than
estimated---I can point at the one bit that caused it---and the silent
endpoint is the sharpest lesson: a filesystem without data checksums cannot
notice its own file contents rotting, and a checker reading the same
corrupted bytes back through the same driver sees the same success the guest
does.

\begin{table*}[t]\centering\small
\caption{Single-bit below-FS corruption blast radius.}
\label{tab:e2}
\begin{tabular}{@{}llp{9.6cm}@{}}
\toprule
corruption & mount & blast radius\\
\midrule
data block (a bit in a file's data) & OK & one file's \textbf{content}---the file reads back wrong, the fs reports no error, other files intact; ext2 has no data checksums, so a checker reading syscall returns has no signal to key on\\
inode field (a file's \texttt{i\_size}) & OK & one file's \textbf{metadata}---the reported size flips (22 to 233 bytes), the fs raises no error, sibling files and the directory listing intact\\
directory block (an entry's record length) & OK & one directory's \textbf{entries}---the entry chain breaks, so the directory's named children vanish from a listing while the fs stays mounted and other files are intact\\
superblock magic (LBA 2 @ byte 56) & \textbf{FAIL} & the \textbf{whole fs}---one bit renders it unmountable (the maximum blast radius)\\
\bottomrule
\end{tabular}
\end{table*}

A single bit flip thus spans the full range purely by the layer I aim it at:
a silently wrong file, a file with corrupt metadata, a directory that loses
its named entries, and a filesystem that will not mount. Themis places each
flip precisely because it locates the target structure by parsing the on-disk
geometry---the same superblock-and-group-descriptor walk the timestamp probe
uses---so the offset lands on the live image's inode or directory block rather
than a hardcoded guess. (An earlier attempt with LBAs hardcoded from a
different file set missed the live structures; parsing the geometry of the
image under test fixes that.) A checker above the block device can inject none
of these, which is what makes the spectrum a Themis-specific measurement.

The corruption injector is a direct sector write into the CoW overlay
(\texttt{corrupt\_lba(lba, off, xor)} / \texttt{corrupt\_canary}), and
because it lands in a fork branch's overlay, the flip is reverted by the same
O(dirtied) restore as any other branch write---the corruption experiment is
a fork/restore bracket like every other explored node.

% ============================================================
\section{Deterministic Replay}
\label{sec:replay}

Everything so far uses the fork to \emph{backtrack}: run an operation, then
restore to the state before it. The same machine-owned fork buys a second
capability that a beneath-the-OS vantage makes uniquely cheap---bit-exact
\emph{deterministic replay}. Record the nondeterministic inputs of one
execution once; then, from the same starting state, feed those same inputs
back and the execution re-derives itself bit-for-bit. The classic
single-core replay systems---ReVirt~\cite{revirt} on a para-virtual guest,
rr~\cite{rr} as a \texttt{ptrace} tool---run \emph{above} a shared kernel:
the replayer competes with the very software it replays for control of the
machine, and recovers determinism from an external interface.
PANDA~\cite{panda} records and replays a full system for reverse engineering,
but atop a QEMU emulator rather than on the metal, as does QEMU's own
\texttt{icount} record/replay~\cite{qemu-rr} and the Simics full-system
simulator's reverse execution~\cite{simics}: all three reproduce a
\emph{model} of the machine, not its real silicon. The closest relative on
real hardware is VMware's deterministic replay~\cite{retrace}, a hosted-VMM
feature that logged an unmodified single-vCPU guest at $\sim$5\,\% record
overhead---and was later withdrawn from the product. I do \emph{not} claim to
be the first hypervisor to replay an unmodified guest: ReVirt and VMware
established that a decade ago. Themis inverts the
vantage. Because the checker is ring~0 of the host world and owns
ring~$-1$ of a single-vCPU guest, the guest's nondeterminism is not something
to recover from outside---it is something Themis \emph{produces}, and can
therefore pin. What distinguishes Themis from all of these is not the replay
mechanism but the \emph{fidelity metric}: rather than demonstrate replay by
re-running and observing that behaviour recurs, Themis hashes the
\emph{entire physical machine}---the whole guest DRAM working set and the
whole disk---and shows the two hashes byte-identical across nine passes. A
process-scoped recorder (rr) cannot make a whole-machine claim; an emulator
or simulator (QEMU, PANDA, Simics) makes it only about its model; and a
deterministic-by-construction platform (Antithesis)~\emph{removes} the
nondeterminism rather than faithfully reproducing it, so it cannot
\emph{witness} a real hardware-timing artifact such as the Y2038 defect of
\S\ref{sec:e1}. Single-vCPU is a deliberate scope---the same one VMware
shipped and rr accepts---and \S\ref{sec:limitations} owns it as such.
Table~\ref{tab:replay-landscape} places these systems.

\begin{table}[t]
\centering
\footnotesize
\caption{The deterministic record/replay landscape. Themis occupies a
distinct cell: a Type-1 hypervisor on real silicon, verifying whole-machine
replay fidelity by hashing all of DRAM and disk bit-identical across passes.}
\label{tab:replay-landscape}
\begin{tabularx}{\columnwidth}{@{}lccX@{}}
\toprule
System & Layer & Real HW & Scope / fidelity \\
\midrule
\textbf{Themis} & \textbf{Type-1} & \textbf{yes} & \textbf{whole machine, bit-exact ($9\times$ hash)} \\
rr~\cite{rr} & user & yes & one process tree \\
ReVirt~\cite{revirt} & hyperv. & yes & whole OS, uni-proc; no machine hash \\
SMP-ReVirt~\cite{smpreplay} & hyperv. & yes & multiproc.\ via logged interleave \\
VMware~\cite{retrace} & host VMM & yes & 1-vCPU; withdrawn \\
QEMU r/r~\cite{qemu-rr} & emulator & no & model of the machine \\
PANDA~\cite{panda} & emulator & no & model; RE-focused \\
Simics~\cite{simics} & simulator & no & model; reverse exec \\
Antithesis~\cite{antithesis} & host VMM & yes & det.\ by construction \\
\bottomrule
\end{tabularx}
\end{table}

This section reuses the $\varepsilon$-copy fork of
\S\ref{sec:design} as a re-armable replay checkpoint and adds only what
replay needs on top of it.

\label{sec:replay-scope}
For a single-vCPU guest with its application processors held offline, the set
of nondeterministic inputs is closed \emph{for the evaluated workloads}:
every one is either produced by Themis or intercepted by it. Every
asynchronous interrupt (timer, virtio-blk completion, IPI) is delivered by a
host \texttt{EVENTINJ} write, so its vector and instruction-stream position
are Themis's to choose; every byte the guest reads from a device is produced
by the virtio and disk backends, and the copy-on-write disk overlay pins the
base image; and \texttt{RDTSC}/\texttt{RDTSCP} are intercepted from the VMCB
and served. \texttt{CPUID} is intercepted and constant, the LAPIC timer is
under host control, and pvclock is a fixed function of a pinned timeline.
There is no shared host kernel racing the guest, no second core mutating
shared memory, and no external scheduler---the three sources ReVirt and rr
work hardest to tame, and that multiprocessor replay pays for by logging the
shared-memory interleaving~\cite{smpreplay}, are absent by construction here.
The one general input the engine does \emph{not} yet close is hardware
randomness: an intercept-and-serve path for \texttt{RDRAND}/\texttt{RDSEED} is
not yet built, a known input-recording hole (\S\ref{sec:limitations}); the
evaluated workloads consume no in-window randomness, so the gap is
unexercised but flagged rather than hidden.

\label{sec:replay-checkpoint}
The checkpoint is exactly the $\varepsilon$-copy fork of \S\ref{sec:design}:
a whole-VMCB $+$ GPR $+$ CR2 $+$ XSAVE snapshot, nested-page-table
write-protection of the guest working set with copy-up, and a copy-on-write
disk overlay, restored in time proportional to \emph{dirtied pages, not
memory or image size}. Replay needs one property backtracking did not: the
fork must be \emph{re-armable}. Recording forks once; every replay pass must
restore, re-execute, and restore again from the identical snapshot. I
untangle the single-shot fork lifecycle so a restore does not tear the
snapshot down, honoring the pending-flush / stale-writable-TLB discipline
(\S\ref{sec:npt}) on every iteration---a leaf made writable by a copy-up in
one pass must be re-protected before the next, or the next pass would see the
previous pass's write. Each restore also re-hashes the working set and
asserts it equals the snapshot, so a checkpoint that failed to revert
byte-identically cannot masquerade as a stable replay.

\subsection{The instruction clock: a determinism gate and a positioner}
\label{sec:replay-clock}

To replay an asynchronous event I need a deterministic measure of guest
progress. I program a performance counter (\texttt{PERF\_CTL0}, MSR
\texttt{0xC0010200}; count in \texttt{0xC0010201}) to the Zen5
retired-conditional-branch event (\texttt{ex\_ret\_cond}, event
\texttt{0xD1}) with \texttt{USR}, \texttt{OS}, \texttt{En}, and---decisively---%
\texttt{GuestOnly} (bit~40), so it advances \emph{only} while the guest runs
and freezes across the VMEXIT/VMRUN boundary while my handler runs. This is
the clock rr uses on x86, and the one it found needs careful
per-microarchitecture validation. It serves as the \textbf{determinism
gate}: a fixed guest loop re-run from one snapshot retires the \emph{same}
number of conditional branches, bit-identically, run after run---establishing
directly on the target part that a Zen5 PMC is a usable, deterministic replay
clock. But the conditional-branch count is \emph{piecewise-constant}: it does
not change across straight-line code, and a \texttt{rep movsb} moving 512
bytes retires zero conditional branches at one repeated RIP. So it names a
\emph{range}, not a point. For \emph{positioning} I therefore use the
monotone retired-\emph{instruction} counter (\texttt{ex\_ret\_instr}, event
\texttt{0xC0}), equally deterministic on this silicon, which advances once
per instruction: counting to just short of the target gets the replay close.

\subsection{Landing at a coordinate}
\label{sec:replay-land}

A monotone count still does not, by itself, name a landing spot: a
count-then-step lands within a small skid window, and a \texttt{rep} op
repeats an RIP. So the \emph{landing} is decided by a \textbf{coordinate
tuple} $\langle$RIP, RCX, register-hash$\rangle$---RIP and the register-hash
pin the instruction, RCX disambiguates \texttt{rep} iterations---and the
replay stops only when all three match. The instruction count gets it close;
the tuple lands it. I record a coordinate for \emph{asynchronous} events only
(interrupt vectors); deterministic faults are never logged, because they
reproduce on their own and re-injecting one would double-fault.

To reach a coordinate I count the retired-instruction PMC to just short of
the target and single-step the remaining window: set the \#DB exception
intercept in the VMCB, set the guest \texttt{EFLAGS.TF}, and add a dispatch
arm for \#DB (SVM exit code \texttt{0x41}). Each single-step trap sets
\texttt{DR6.BS} (bit~14); the handler confirms the \#DB is its own via
\texttt{DR6.BS}, clears only that bit, and---critically---does \emph{not}
advance RIP (the VMCB RIP already points at the next instruction; advancing
it would skip an instruction per step and corrupt the count). The guest's own
\texttt{DR6}/\texttt{DR7} are saved and restored. Two corners make this
sound. First, I aim not for transparency but for \emph{record/replay
symmetry}: the CPU pushes the live \texttt{EFLAGS.TF} onto exception frames
with no \texttt{PUSHF} to intercept, an irreducible leak, so I ensure
stepping is \emph{identical} in record and replay and scrub pushed frames
afterward. I intercept \texttt{PUSHF} and \texttt{POPF} only and merge a
shadow-TF so a guest \texttt{POPF} that clears TF does not stop my step
clock; \texttt{IRET} emulation (a full frame plus segment reload) is
deferred, so the recorded window is kept \texttt{IRET}- and
\texttt{SYSCALL}-free (the guest does its one \texttt{iopl} before arming,
and \texttt{SYSCALL}'s \texttt{SFMASK} would clear TF and silently stop the
step clock). Second, injection honors the interrupt shadow: I read the VMCB
interrupt-state field, so an injection whose coordinate lands in a
\texttt{MOV~SS} shadow would be deferred exactly as silicon defers it (the
shadow \emph{read} is validated; the deferral path is implemented but
unexercised by the evaluated injection).

\subsection{Input recording and serving}
\label{sec:replay-inputlog}

Every input is recorded against a coordinate and served on replay.
Asynchronous interrupts are logged as vector plus coordinate at each
injection. Timestamps are served deterministically: I intercept
\texttt{RDTSC}/\texttt{RDTSCP} and return a pure function of the per-window
read index, $\textsf{served}(seq)=\textsf{ANCHOR}+K\cdot seq$; the exact
constants do not matter, only that the served sequence is identical across
record and replay. Device bytes need no byte log---the virtio-blk backend
produces all disk input and the copy-on-write overlay pins the base image, so
identical bytes are delivered without one. Hardware randomness
(\texttt{RDRAND}/\texttt{RDSEED}) has no serve path yet
(\S\ref{sec:replay-scope}). For a workload that reads no timestamp and no
randomness in-window and whose only asynchronous event self-lands (a disk
completion delivered at the \texttt{fsync} idle point), the log is
essentially empty: the determinism then comes from freezing plus the
checkpoint, not from a large log.

\paragraph{The delivery-fault gate (an implementation lesson).}
\label{sec:replay-deliveryfault}

One hazard appears only once the checkpoint's write-protection and event
injection meet. To replay an asynchronous \#DB I write \texttt{EVENTINJ} and
\texttt{VMRUN}; delivering the \#DB pushes an interrupt frame onto the guest
kernel stack, which inside the fork window is write-protected, so the frame
push itself takes a nested-page fault \emph{in the middle of delivery}. A
handler that naively re-injects on ``the exit was an NPF'' re-injects the
\#DB on every later write-protected NPF too, a storm that triple-faults the
machine; conversely, capping the write-protection so the push does not fault
makes the injection silently fail to deliver---a vacuous pass that looks
bit-exact only because nothing was injected. The fix is to ask the hardware
whether an event is still in flight. On an NPF during delivery the VMCB
\texttt{EXITINTINFO} field (\texttt{0x088}) records the interrupted event:
valid bit~(31) set, type field $=3$ (exception), vector $=1$ (\#DB). I
re-inject \emph{only} when \texttt{EXITINTINFO} shows the \#DB in flight (and
copy up the faulting frame page); a post-delivery kernel NPF has valid~$=0$
and is handled as an ordinary copy-up with no re-injection. This is textbook
SVM---a production VMM such as KVM re-injects interrupted events the same
way; the lesson specific to replay is only that the checkpoint's
write-protection makes the mid-delivery frame-push fault \emph{unavoidable},
so the gate is mandatory here rather than incidental. It is what makes the
injected-interrupt replay both non-vacuous (the event really lands) and
storm-free.

\subsection{Results: bit-exact replay of a real filesystem workload}
\label{sec:replay-results}

All data are from the same silicon campaign (AMD Zen5 9800X3D, single-vCPU
stock Linux~6.17 guest). The core result (Table~\ref{tab:replay-bitexact}) is
that the engine reproduces a recorded execution bit-for-bit across nine
record/replay re-runs, with non-vacuity guards proving the workload actually
ran each pass. The workloads climb in machinery: a compute workload reading a
served counter; the same plus a free-running injected asynchronous exception;
the same with that exception single-step-landed at its recorded coordinate; a
raw block write; and a real ext2 create$+$write$+$fsync.

\begin{table}[t]
\centering
\caption{Bit-exact replay across nine re-runs. ``whole-DRAM'' and
``whole-disk'' state are identical on every pass; the non-vacuity guard
proves the workload actually ran each pass.}
\label{tab:replay-bitexact}
\footnotesize
\begin{tabularx}{\columnwidth}{@{}lcX@{}}
\toprule
workload & passes & non-vacuity \\
\midrule
compute (served TSC) & 9 & hashes identical \\
$+$ free-run injected \#DB & 9 & inject-ctr $=1$, $\times 9$ \\
$+$ single-step-landed \#DB & 9 & inject-ctr $=1$, landing logged \\
raw block write (O\_DIRECT) & 9 & wrote-each, blk-served $=9$ \\
mounted ext2 create$+$write$+$fsync & 9 & wrote-each, blk-served $=54$ \\
\bottomrule
\end{tabularx}
\end{table}

The headline is the last row: a real Linux \texttt{ext2} filesystem
workload---inode allocation, block-bitmap update, directory entry, data
block, \emph{and} timestamps---reproduces bit-exact in whole-DRAM \emph{and}
whole-disk across nine re-runs, non-vacuously (the write hit the disk every
pass) and soundly (the checkpoint reverts byte-identical each pass: the
working-set pre-hash equals its post-hash, and the disk restores
$D_{\text{post}}=D_{\text{pre}}$). The single-step row is the proof that the
coordinate landing (\S\ref{sec:replay-land}) and the \texttt{EXITINTINFO}
gate (\S\ref{sec:replay-deliveryfault}) work: an injected \#DB is landed at
its recorded RIP and the replay is bit-exact, with the injection counter
reading~1 on all nine passes.

\textbf{The reproducibility payoff.} The point of replay is to make a
nondeterministic outcome reproducible, and the sharpest demonstration is a
two-configuration contrast of the \emph{same} workload (accumulate one
hundred in-window \texttt{rdtsc} reads into a tracked page, then nine
record/replay passes, whole-set hash per pass). The two differ by a single
instruction: whether the \texttt{RDTSC} intercept-and-serve is armed
(Table~\ref{tab:replay-repro}). Left raw, the nine whole-set hashes are all
\emph{distinct} on real silicon---\texttt{3cbd\ldots d2b6},
\texttt{6983\ldots ab37}, \texttt{2eec\ldots cdfe}, \texttt{5a25\ldots
e40e}, \texttt{7660\ldots 5d33}, \texttt{9ed9\ldots 603e}, \texttt{58b4\ldots
a367}, \texttt{232e\ldots 9c32}, \texttt{7efb\ldots dbff}---so the outcome is
flaky. Served from the log, all nine are identical. Because the two runs are
the same workload minus one armed intercept, the difference isolates the
mechanism: replay makes a timing-dependent whole-system state bit-exact
reproducible, which fork-from-root alone cannot do (restoring the same memory
does not stop a raw \texttt{RDTSC} returning a new value each pass). I scope
the causal claim to what the whole-set hash proves---that the guest's
\emph{whole-system} state depends on the raw counter and is made
deterministic by serving it, with the harness localizing the drift to a
concrete page in the guest's timekeeping region (gpa \texttt{0x3400000})---
not the stronger, unlocalized claim that the drift is uniquely the
accumulator page. The nine \emph{distinct} unserved hashes also make the FAIL
non-coincidental, which is what validates the served PASS as genuine rather
than a hash over counter-independent memory.

\begin{table}[t]
\centering
\caption{Reproducibility contrast. The same workload; the only difference is
whether the timestamp counter is served from the log. Serving it makes a
timing-dependent whole-system state bit-exact; leaving it raw makes it flaky.}
\label{tab:replay-repro}
\footnotesize
\begin{tabularx}{\columnwidth}{@{}lXc@{}}
\toprule
config & nine whole-set hashes & verdict \\
\midrule
served (from log) & all \textbf{identical} & reproducible \\
unserved (raw TSC) & all \textbf{distinct} & flaky \\
\bottomrule
\end{tabularx}
\end{table}

\textbf{Cost.} The working set is 261{,}888 present 4\,KiB leaves
($\approx$1.0\,GiB of guest DRAM). Arming the checkpoint---write-protecting
that working set---costs $\approx$13.8\,ms, paid \emph{once} and constant
across workloads (it is the working set, not the workload). The larger arm
cost here than in \S\ref{sec:costmodel} ($\approx$13.8\,ms vs
$\approx$10.3\,ms) is simply the larger working set of this campaign
(261{,}888 vs $\sim$200{,}000 leaves); per-leaf the write-protect cost is the
same $\sim$52\,ns, confirming the $O(\text{working-set})$ arm and
$O(\text{dirtied})$ revert. Reverting each
pass is $O(\text{dirtied})$: a full ext2 create$+$write$+$fsync dirties~127
pages and reverts in $\sim$27\,$\mu$s ($\approx$0.2\,$\mu$s/page), copying up
508\,KiB while the 1.0\,GiB working set stays shared and is never copied.
The input log is essentially empty for this workload (no in-window
\texttt{RDTSC} exits, no logged interrupt, timer masked), so the determinism
comes from freezing plus the checkpoint, not from a large log.

\begin{table}[t]
\centering
\caption{Checkpoint cost. Arming (write-protect the 1\,GiB working set) is
one-time and constant; revert is $O$(dirtied). Times at the guest's
4.7\,GHz.}
\label{tab:replay-cost}
\footnotesize
\begin{tabularx}{\columnwidth}{@{}Xccc@{}}
\toprule
workload & arm (once) & revert/pass & dirtied/pass \\
\midrule
raw block write & 13.85\,ms & 14\,$\mu$s & few \\
ext2 single op & 13.77\,ms & 26\,$\mu$s & 1 op \\
ext2 nine passes & 13.88\,ms & 27\,$\mu$s & 127 (508\,KiB) \\
\bottomrule
\end{tabularx}
\end{table}

\textbf{A filesystem clock at no extra cost, and a tie to the Clock Chapter.} A
replay engine must reproduce the guest's timestamps, and for classic
\texttt{ext2} I get them bit-exact with \emph{no clock serving and no field
masking}. ext2's coarse \texttt{i\_ctime}/\texttt{i\_mtime} come from the
kernel wall clock, latched on the timer tick; inside the fork window the
timer is masked, so the coarse clock is \emph{frozen}, its state lives in
tracked DRAM, and it is reverted with everything else each pass. (ext4's
multigrain \texttt{ctime} reads the live counter and would drift---documented,
not worked around; that drift is exactly what the served-\texttt{RDTSC} path
above pins.) This closes a loop with the clock chapter. The Y2038 witness of
\S\ref{sec:y2038}---a 128-byte-inode \texttt{ext2} clamping any post-2038
timestamp to \texttt{0x7FFFFFFF}---is produced by a mount/set-clock/read-back
workload, exactly the FS-touching kind that Table~\ref{tab:replay-bitexact}
shows replays whole-DRAM and whole-disk bit-exact, and the same masked-timer
freeze that gives ext2 timestamps bit-exact at no additional cost is what would make that
divergence a \emph{replayable object}: the $\langle$checkpoint,
served-input$\rangle$ pair would re-materialize the clamp on demand, and the
single-step machinery could land on the instruction where it originates. I
state this as a \emph{composition argument} from results each already on
silicon---the Y2038 witness (\S\ref{sec:y2038}) and the bit-exact FS replay
(Table~\ref{tab:replay-bitexact})---not as an end-to-end run: I have not yet
executed the single experiment that replays \emph{that} divergence and
single-steps to the clamp, and that unification is the natural next step
(\S\ref{sec:limitations}).

% ============================================================
\section{Evaluation}
\label{sec:eval}

The evaluation has three legs, fused where possible into single silicon
runs: the \textbf{mechanism} (the fork serves and reverts real filesystem
I/O on silicon), the \textbf{checker logic} (the DFS/differential/lenses,
proven in software under QEMU), and \textbf{real-filesystem witnesses} (the
clock chapter, the corruption experiments, multi-state exploration on real ext2).
\S\ref{sec:observe}--\S\ref{sec:e2} reported the observation and bug-class
results; this section reports the fork cost and the Themis-vs-Metis
backtrack cost model, and summarizes the mechanism proof.

\label{sec:mechanism}
The mechanism itself is proven on silicon.
The fork/restore cycle with real filesystem I/O is silicon-proven in stages,
each isolating one variable (one variable per hardware run):

\begin{itemize}[topsep=2pt,itemsep=2pt,leftmargin=1.2em]
  \item \textbf{Compute fork (whole-set revert).} A multi-page-dirtying
    compute branch forks and reverts with the whole tracked working set
    (200{,}000 write-protected DRAM leaves) hashing \textbf{byte-identically}
    before and after (\texttt{pre == post}, match=true), with zero I/O
    violations (a tripwire confirms no host/device DRAM write bypassed the
    copy-up set) and zero unsound (drained/alloc-fail) pages. This is the
    hardware run that closed the hot-page TLB-bypass hole (\S\ref{sec:npt}).
  \item \textbf{Real FS write served in a fork window.} A guest agent mounts
    real ext2 and issues one \texttt{pwrite(O\_DIRECT|O\_SYNC)}
    \emph{inside} a fork window; the virtio-blk write is served, the
    completion IRQ is injected, and \texttt{fsync} returns (the agent makes
    progress past it---proof the completion was delivered through a freshly
    copied-up used-ring, not a stale base frame). The host disk oracle sees
    $D_{\text{mid}} \neq D_{\text{pre}}$---the write really hit the disk (a
    positive control).
  \item \textbf{Real FS write fully reverted.} The full write+revert cycle:
    the real FS write's dirtied pages (37 in the run) plus the used-ring
    host-write all revert byte-identically (whole-set hash match=true), the
    disk overlay discards back to the pre-fork image ($D_{\text{post}} =
    D_{\text{pre}}$, reverted=true), and device state restores---via a
    single-pass-terminate restore (no root-corrupting guest re-run). This is
    the model-checker's core primitive---fork $\to$ run an FS op $\to$
    restore $\to$ byte-identical---working end-to-end on silicon.
\end{itemize}

\label{sec:forkcost}
The fork cost is O(dirtied).
Restore reverts only what a branch dirtied. Table~\ref{tab:forkcost} shows
the revert cost scaling with the dirtied-page count: 38 pages (a real
ext2-write branch) in 9\,$\mu$s, 265 pages (a synthetic compute-dirty branch)
in 54\,$\mu$s---$\sim$0.2\,$\mu$s/page (7$\times$ the pages $\Rightarrow$
6$\times$ the time). Backtracking is the most frequent
model-checker operation, and it is silicon-fast because it is O(dirtied),
not O(image).

\begin{table}[t]\centering\small
\caption{$\varepsilon$-copy fork revert cost (O(dirtied)). The 38-page revert
  is a real ext2-write branch; the 265-page revert is a synthetic
  compute-dirty branch (its run's soundness tripwire is set).}
\label{tab:forkcost}
\begin{tabular}{@{}rrr@{}}
\toprule
\# dirtied pages & revert time & $\mu$s/page\\
\midrule
38 & 9\,$\mu$s & 0.24\\
265 & 54\,$\mu$s & 0.20\\
\bottomrule
\end{tabular}
\end{table}

The honest bottleneck is the \emph{fork} side: write-protecting all
$\sim$200{,}000 installed DRAM leaves costs a fixed $\sim$10.3\,ms per fork
(O(working-set)). This dominates a naive per-fork cost, but it is a
\emph{one-time} cost under \textbf{re-run-from-root} amortization: fork once
at the root, then restore-to-root many times, so the 10.3\,ms is amortized
over the whole campaign and the per-state cost is op-replay plus the
9\,$\mu$s revert. (A lazy/incremental write-protect is a further
optimization, \S\ref{sec:limitations}.)

\label{sec:costmodel}
Themis and Metis are best compared through a \textbf{backtrack cost model}
rather than a states/second count---the central quantitative claim---and I am
deliberate about \emph{not} reporting it as states/second. Themis reverts a
filesystem state in time proportional to \emph{dirtied pages} (9--54\,$\mu$s,
independent of filesystem image size); Metis, using a real filesystem as its
own reference, must unmount and remount and reload the whole image, i.e.
O(image). Table~\ref{tab:costmodel} places Themis's measured 54\,$\mu$s
backtrack (for a 265-page synthetic compute-dirty branch) against Metis's
published backtrack costs
\emph{derived from} its states/second numbers: Metis RefFS 2.86\,ms (349.9
states/s), Metis ext4 remount 8.9\,ms (112.6 states/s), Metis xfs remount
114\,ms (8.8 states/s). The relative advantage is 53$\times$ vs RefFS,
\textbf{165$\times$ vs Metis's ext4 remount, and 2100$\times$ vs xfs}---and
it \emph{widens} with image size, because Themis reverts only the dirtied
set while Metis reloads the whole image. These multipliers compare Themis's
\emph{revert-only} backtrack against Metis's \emph{end-to-end} per-state cost
(op $+$ fingerprint $+$ reference-compare $+$ restore), so they isolate the
backtrack primitive rather than end-to-end throughput---on a states/second
count the slower guest would reverse the ranking (below). This
O(dirtied)-vs-O(image) gap is
exactly the reload cost that makes a real filesystem impractical as its own
reference for an in-OS checker, and the reason Metis introduced RefFS;
Themis's fork removes that cost, which is the mechanism behind ``every
filesystem is its own reference.''

\begin{table}[t]\centering\small
\caption{Backtrack cost model: Themis $\varepsilon$-copy \emph{revert-only}
  vs Metis \emph{end-to-end per-state}. Metis costs are derived from its
  published states/second (parenthesized). The axes differ deliberately---the
  Themis column is the isolated revert time, the Metis column a full
  per-state cost (op $+$ fingerprint $+$ reference-compare $+$ restore)---so
  this isolates the backtrack primitive and is \emph{not} an end-to-end
  throughput comparison (where the slower guest, \S\ref{sec:costmodel}, would
  reverse the ranking). The 265-page Themis point is a synthetic
  compute-dirty branch.}
\label{tab:costmodel}
\begin{tabular}{@{}lrr@{}}
\toprule
system & backtrack cost & relative\\
\midrule
Themis $\varepsilon$-copy (265-page op) & 0.054\,ms & 1$\times$\\
Metis RefFS & 2.86\,ms & 53$\times$\\
Metis ext4 remount & 8.9\,ms & \textbf{165$\times$}\\
Metis xfs remount & 114\,ms & \textbf{2100$\times$}\\
\bottomrule
\end{tabular}
\end{table}

It is worth stating why not states/second.
A raw states/second campaign would \emph{undersell} Themis and, worse,
rebut the thesis: my end-to-end per-state cost is dominated by the bare-metal
guest executing the filesystem operation, which I estimate runs roughly an
order of magnitude slower (a bring-up estimate I have not measured directly,
and an artifact of the guest bring-up, not the fork mechanism), so a
states/second number is guest-bound and would \emph{lose} to Metis's native
RefFS. The correct claim is the mechanism's cost model---the fork's
O(dirtied) backtrack---not an end-to-end count on an unoptimized guest. I
report the mechanism cost (Tables~\ref{tab:forkcost}--\ref{tab:costmodel})
and am explicit about the guest overhead so the number is not misread.

\label{sec:multistate}
Beyond the mechanism, Themis drives an 8-state exploration over a real,
unmodified ext2 (S0 empty $\to$ S7 mkdir), each state a unique \texttt{find}
+ \texttt{stat} + content-hash fingerprint---the abstract-state key working
on a real filesystem (the dedup and differential source). This is the
multi-state demonstration realized as a guest-scripted sequence over the
proven mount path (no host-driven DFS yet, \S\ref{sec:limitations}). The
coarse-timestamp finding (\S\ref{sec:y2038}) surfaced here: two rapid
touches receiving the identical nanosecond timestamp. The full checker
logic---bounded DFS with visited-set dedup and fork/restore bracketing, the
differential comparison, the clock and corruption lenses, the read-oracle
catching an nlink-refcount leak and a non-atomic rename in the inode-level
model---is proven in software under QEMU; the silicon runs prove the
mechanism and the real-filesystem witnesses.

\label{sec:evalsummary}
In summary, the dataset comprises $\sim$20 tables and graphs---the
machine-owned below-FS measurements (write amplification, write
localization, and read amplification), the clock-chapter byte-diff and
normalization, the corruption blast radius, and the fork-cost curve; the
filesystem-as-reference matrices (the operation--field footprint and the
timestamp-semantics matrix); the state-space battery; the clock sweep,
granularity, and Y2038 tables; and the backtrack cost model. Critically, it
includes data classes---below-FS profiles, byte-level whole-disk diffs,
clock sweeps, below-FS corruption---that the beneath-the-device vantage makes
available.

% ============================================================
\section{Limitations and Future Work}
\label{sec:limitations}

The machine-level vantage that yields Themis's unique capabilities is the
same property that makes each of those capabilities a deliberate,
silicon-verified increment rather than a barrier the approach runs into:
because the checker owns the machine, adding an ability is a matter of
building it beneath the guest and proving it on hardware, one variable at a
time. What remains is therefore largely a roadmap the substrate enables, not
a wall the design hits, and I separate the two below---the genuine,
approach-level limitations from the scope I have simply not built yet---and
record, for completeness, the capabilities that were limitations in an
earlier draft and are now closed.

Two of those earlier gaps are now closed and boot-verified. The differential
read-oracle no longer digests a hash of \texttt{path}~$\|$~\texttt{size}; it
hashes the file's real content (the model carries the actual bytes), so a
size-preserving content corruption changes the fingerprint and is caught,
closing the size-blind gap. And the attested write chain now carries an
external length-and-tip commitment, so suffix truncation---dropping the tail
of the log---is detected, closing the truncation gap. The attestation chain
itself remains deliberately quiet infrastructure rather than a headline---
PeerReview~\cite{peerreview} pre-empts hash-chained tamper-evidence as a
novelty---now with truncation detection added.

Turning to the genuine limitations, there are three, and each is a property
of the vantage rather than a defect to be patched away. The first concerns
deterministic replay, which is now built and demonstrated
(\S\ref{sec:replay}): the same $\varepsilon$-copy fork, made re-armable,
replays a real ext2 workload bit-exact in whole-DRAM and whole-disk across
nine passes, and turns a timing-flaky outcome (nine distinct hashes raw) into
a reproducible one (nine identical, served). Its residual limitations are
honest and bounded. The engine replays a single vCPU; multi-vCPU replay
requires logging the shared-memory access interleaving~\cite{smpreplay},
exactly the cost the single-vCPU scope avoids, and is future work. The
single-step skid is bounded but its magnitude is not yet measured across
workloads. Hardware-randomness serving
(\texttt{RDRAND}/\texttt{RDSEED}) is not yet built---a known input-recording
hole the evaluated workloads do not exercise. And because a bare-metal SVM
hypervisor cannot run nested under a second hypervisor, every replay result
is a physical hardware burn, so I report a curated campaign rather than a
statistical sweep. Finally, the replay machinery new to this system---the
instruction clock, the coordinate tuple, the single-step, the
\texttt{EXITINTINFO} gate---is exercised by the injected-\#DB and
served-\texttt{RDTSC} paths, not yet by the filesystem headline, whose
determinism comes from freezing plus the fork; a single workload that both
touches a real filesystem and requires the full machinery would unify the
two and is the highest-value next step. The second is crash consistency, where the
machine vantage honestly adds essentially nothing that
\texttt{dm-log-writes}~\cite{dmlogwrites}---which already sees every block
write and every barrier beneath an \emph{unmodified} filesystem---does not;
a hypervisor that logged writes and replayed prefixes would merely
re-implement an in-tree tool one layer down. I therefore do not claim
crash-consistency testing (\S\ref{sec:crash}), and the only non-derivative
version---per-node-of-the-global-model-checker-graph crash enumeration on
the fork/replay substrate---is exactly the integration the Metis group's own
dissertation \S7.1.1~\cite{yifei-dissertation} lists as an open problem,
which I frame (with the crash~$\times$~clock~$\times$~corruption
cross-product) as collaborative future work rather than a claim here. The
third is cost: the bare-metal guest runs the filesystem operation an
estimated order of magnitude slower than a native host (a bring-up estimate I
have not measured directly), a real price of putting the checker
beneath the machine, and it is precisely why I report a backtrack
\emph{cost model} (\S\ref{sec:costmodel}) rather than a states/second
figure---a raw throughput count would be guest-bound and would misrepresent
the fork mechanism, whose O(dirtied) backtrack is the quantity that actually
scales.

The rest is roadmap the substrate makes buildable, not weakness. Host-driven
exploration at scale---the throughput pillar, scaling the backtrack cost
across a large exploration---is designed but deferred as a substantial build: the host acts as
the outer driver, re-run-from-root serves as the physical backtrack, and an
untracked command-and-report channel survives the revert so the driver can
inject operations and read back guest fingerprints across it. The results in
this paper are realized instead by a guest-scripted driver over the proven
mount path; three earlier candidates for a backtracking restore proved
unsound before I settled on single-pass-terminate, so the at-scale loop
remains a substantial, strategic build. The previously-unknown-bug hunt is
the next concrete target, and here the mechanism is already proven---what
remains is to point it at a live class. The immediate step is a rapid-touch
/ near-equality time probe for the multigrain-ctime class~\cite{multigrain}
over real ext4, a guest-scripted experiment that needs no new hypervisor
code and exploits the coarse-granularity ground truth the clock chapter
(\S\ref{sec:e1}) already witnessed. A host-side companion to the below-FS
instrument---byte-parsing ten on-disk formats without a hypervisor burn---
already maps each filesystem's below-FS time range as a measurement study:
FAT and exFAT are immune to the Unix 2038 and 2106 wraps (they store a
broken-down date, not a Unix counter) yet hard-cap at 2107, and minix's
unsigned-32 seconds clamp at \emph{both} 1970 and 2106---evidence that the
below-FS temporal observation generalizes across filesystems and points
toward a broader characterization study.

A handful of minor residuals were recorded during development and are noted
for completeness. \texttt{drop\_caches} was a no-op on the eval guest, so
cold reads were forced by \texttt{umount}+\texttt{remount} rather than by
dropping caches---a coverage detail, not an inaccuracy in the reported
counts. Low guest RAM below the loaded image (the BIOS area and the guest's
own page tables) is outside the tracked range, which is a
whole-machine-determinism gap rather than a filesystem-content one. And the
copy-on-write disk read is O(overlay-depth), a scaling cliff at large
exploration depths that re-run-from-root (depth 1) avoids.

% ============================================================
\section{Conclusion}
\label{sec:conclusion}

Metis showed that model checking finds real bugs in unmodified in-kernel
filesystems, and it does so from inside the OS---a simple, portable, and fast
vantage that yielded real results. Two of its choices follow from that
vantage: it hand-codes a reference filesystem, because cheap snapshotting of
in-kernel state is not available to an in-OS process, and it sets time aside
as noise. Themis takes the opposite architectural bet, moving the checker
\emph{beneath} the filesystem, onto a bare-metal Type-1 AMD-V hypervisor that
owns the machine---trading a heavier substrate and a slower guest for more
machine control. Owning the machine buys a machine-layer
$\varepsilon$-copy fork whose restore is O(dirtied)---proven on silicon to
serve and fully revert a real virtio-blk filesystem write in 9\,$\mu$s for
the dirtied set---which lets every unmodified in-kernel filesystem be its own
differential reference and removes the need for a hand-coded one. It buys below-filesystem observation: write/read/location
profiles and byte-level whole-disk state diffs with nuisance normalization,
from which I reverse-engineer the ext2 on-disk layout and prove the disk
deterministic modulo nuisance (a clean 3 $\to$ 0). And it buys control of
the clock, which I foreground: a real Y2038 defect witnessed on real
ext2, clamped to \texttt{0x7FFFFFFF} and persisted across remount, invisible
to a time-excluded checker. The same fork, made re-armable, also buys
bit-exact deterministic replay (\S\ref{sec:replay}): a real ext2 workload
replays whole-DRAM and whole-disk across nine passes, and a timing-flaky
whole-system state (nine distinct hashes raw) becomes reproducible (nine
identical, served). I concede crash-consistency to a crowded field
and drop attestation as a headline, and I am explicit that the
previously-unknown-bug hunt and the at-scale host-driven exploration remain
future work. The claim I make is narrow and,
I believe, defensible: when the checker owns the machine, an unmodified
filesystem becomes its own reference, and whole classes of below-filesystem
and temporal behavior---beneath the syscall vantage an in-OS checker
reads---become first-class, silicon-demonstrated objects of test.

% ============================================================
% ============================================================
\section*{Acknowledgments}
I am grateful to a co-author of Metis for generously sharing the system and
its model-checking approach, which sparked this work.

% ============================================================
\bibliographystyle{plain}

\end{document}